\documentclass[a4paper,10pt]{article}
\pdfoutput=1 

\usepackage{jheppub} 

\usepackage[T1]{fontenc} 

\usepackage{graphicx}
\usepackage{verbatim,epsfig}
\usepackage{epstopdf}
\usepackage[utf8]{inputenc}
\usepackage[colorlinks=true,linkcolor=blue,citecolor=blue]{hyperref}
  \usepackage{bm}
   \usepackage{amsmath}
    \usepackage{amssymb}
     \usepackage{pifont}

\newcommand{\nn}{\nonumber}

\newcommand{\Tr}{\mathrm{Tr}}

\newcommand{\ot}{\leftarrow}

\renewcommand{\(}{\left(}
\renewcommand{\)}{\right)}
\renewcommand{\[}{\left[}
\renewcommand{\]}{\right]}

\title{Structure of rapidity divergences in soft factors}

\author{Alexey Vladimirov}

\affiliation{Institut f\"ur Theoretische Physik, Universit\"at Regensburg,\\
D-93040 Regensburg, Germany}

\emailAdd{alexey.vladimirov@physik.uni-regensburg.de}

\abstract{
We discuss the structure of rapidity divergences that are presented in the soft factors of transverse momentum dependent (TMD) factorization theorems. To provide the discussion on the most general level we consider soft factors for multi-parton scattering. We show that the rapidity divergences are result of the gluon exchanges with the distant transverse plane, and are structurally equivalent to the ultraviolet divergences. It allows to formulate and to prove the renormalization theorem for rapidity divergences. The proof is made with the help the conformal transformation which maps rapidity divergences to ultraviolet divergences. The theorem is the systematic form of the factorization of rapidity divergences, which is required for the definition of TMD parton distributions. In particular, the definition of multi parton distributions is presented. The equivalence of ultraviolet and rapidity divergences leads to the exact relation between soft and rapidity anomalous dimensions. Using this relation we derive the rapidity anomalous dimension at the three-loop order.
}

\begin{document}
\maketitle
\flushbottom

\section{Introduction}

Soft factors are the inherent part of the modern factorization theorems, and have the common structure of vacuum matrix elements of a product of Wilson lines. The geometrical configuration of a soft factor reflects the classical picture of the scattering. In this way, the massless initial- and final- state partons represent themselves as a half-infinite lightlike Wilson lines, and give rise to the variety of divergences. Apart of usual collinear and ultraviolet (UV) divergences, they produce rapidity divergences. The latter is a subject of special treatment and factorization procedure. Many aspects of rapidity divergences are still unstudied. In this paper, we present  the study of the rapidity divergences, and their connection to the geometry of soft factor. We demonstrate that the rapidity divergences are related to particular spatial configurations, and formulate the requirements on the structure of soft factors that guaranty the factorization of rapidity divergences. It allows us to formulate and prove the renormalization theorem for rapidity divergences.

To make an introduction to the problem, and demonstrate its practical importance, let us recall the transverse momentum dependent (TMD) factorization theorem, where the rapidity divergences and their factorization play one of the central roles. The TMD factorization theorem describes such processes as Drell-Yan (DY) and semi-inclusive-deep-inelastic-scattering (SIDIS) in the regime of low transverse momentum $q_T$. Within the TMD factorization the expression for the hadron tensor takes the form (see e.g. \cite{Becher:2007ty,Becher:2010tm,Collins:2011zzd,Collins:2011ca,GarciaEchevarria:2011rb})
\begin{eqnarray}\label{intro:W=HFSF}
W_{\text{TMD}}=H\otimes \[\bar F(\delta^-) S(\delta^-,\delta^+) F(\delta^+)\]+\mathcal{O}\(\frac{q_T}{Q}\),
\end{eqnarray}
where $H$ is the Wilson coefficient for the hard-collinear matching, $F$ and $\bar F$ are hadron matrix elements of collinear and anti-collinear fields, and $S$ is the TMD soft factor. Here, the argument $\delta$ represents a regulator for rapidity divergences associated with particular hadron. The rapidity divergences cancel in the product of factors, so the expression (\ref{intro:W=HFSF}) is finite. However, the factorization formula (\ref{intro:W=HFSF}) is not practical, because it does not define a measurable parton density. The main difficulty is caused by the soft factor which mixes the rapidity divergences of both hadrons. To finalize the factorization and to define universal TMD parton distribution one has to perform the factorization of rapidity divergence in the soft factor.

There are several approaches to formulate the factorization of rapidity divergences for the TMD soft factor. The differences among approaches are originated from the differences in regularization schemes. It appears to be difficult to find a commonly convenient regularization for the rapidity divergences, since they are insensitive to the dimensional regularization \cite{Collins:1992tv}. Nowadays, there are three most popular approaches to the factorization of rapidity divergences. \textit{(i)} Explicit evaluation of the soft factor and the manual split of divergent contributions \cite{Collins:2011zzd,Collins:2011ca,GarciaEchevarria:2011rb,Echevarria:2012js}. In this case, the soft factor takes the form of the product of divergent terms, like ${S(\delta^-,\delta^+)=\sqrt{S(\delta^-;\zeta)}\sqrt{S(\delta^+;\zeta)}}$. \textit{(ii)} Formulation of a scaleless regularization for rapidity divergences in which the soft factor is unity at all orders of perturbation theory (e.g. the analytical regularization \cite{Smirnov:1997gx}). The effect of factorization arises via an anomalous-like contribution, aka collinear anomaly  \cite{Becher:2010tm,Becher:2012yn}. \textit{(iii)} Subtraction of the rapidity divergences at the symmetric point by the renormalization procedure similar to the UV renormalization \cite{Chiu:2011qc,Chiu:2012ir}.  All three schemes have been checked by explicit next-to-next-to-leading order (NNLO) calculations (see \cite{Echevarria:2015byo,Echevarria:2016scs} for \textit{(i)}, \cite{Gehrmann:2012ze,Gehrmann:2014yya} for \textit{(ii)} and \cite{Luebbert:2016itl,Li:2016axz} for \textit{(iii)}). The results agree with each other. 

In fact, all these schemes imply that the logarithm of soft factor is linear in the rapidity divergence, i.e. $\ln S(\delta^-,\delta^+)\sim \ln(\delta^+\delta^-)$ (here, the divergences are represented by $\ln \delta$). This statement automatically leads to the factorization of rapidity divergences and to the equivalence of all approaches. The linearity in the rapidity divergences seems natural. Indeed, the structure of the exponentiated diagrams for the TMD soft factor is rather simple and gives some intuition how the cancellation of higher-order divergences takes place. This intuition is also supported by NNLO calculation. However, the factorization procedure is not proven, to our best knowledge. The absence of any proof conceptually prevents the extension of factorization to more difficult processes such as multi-parton scattering \cite{Diehl:2011yj,Manohar:2012jr,Diehl:2015bca}, or processes with a richer final state that involve complicated soft factors (see e.g.\cite{Stewart:2010tn,Jouttenus:2011wh}).

To access the problem of rapidity divergences on a more general level, we study the multi-parton scattering (MPS) factorization and its soft factor. Structurally, the factorized MPS hadron tensor repeats the DY hadron tensor (\ref{intro:W=HFSF}), but obtains a non-trivial color structure,
\begin{eqnarray}\label{intro:W=HFSF2}
W_{\text{MPS}}=\sum_{i=1}^N H_i\otimes \[\bar F_{a_1...a_N}(\delta^-) S^{a_1...a_N,b_1...b_N}(\delta^-,\delta^+) F_{b_1...b_N}(\delta^+)\]+\mathcal{O}\(\frac{q_T}{Q}\),
\end{eqnarray}
where $N/2$ is the number of partons involved in the MPS. The MPS factorization is a direct generalization of TMD ($N=2$), and double-parton scattering ($N=4$) cases. Practically, the MPS is not that important, since it is only the one of many channels contributing to the multi-particle production reaction. However, theoretically, it is very interesting, and allows to look at the problem of rapidity divergences from a new side. In particular, it clearly shows that the rapidity divergences are associated with planes rather than with vectors, which is typical assumption. Therefore, the MPS soft factor has only two rapidity divergences, although it is a composition of many Wilson lines. To our best knowledge, the MPS soft factor has not been studied. Therefore, we start the paper from the presentation of details on the structure of MPS soft factor in sec.\ref{sec:MPS_SF}. It includes the presentation of the all-order color-structure and NNLO expression in sec.\ref{sec:color}.

The association of the rapidity divergences with the planes has far going consequences. First, it gives the simple and intuitive geometrical criterion of non-overlapping rapidity divergences, namely, the corresponding planes should not intersect. Second, it allows the conversion of rapidity divergences to UV divergences by a conformal transformation. In sec.\ref{sec:Cnn} we construct the conformal transformation which maps the distant transverse plane to a point, and demonstrate the transition of divergences. The equivalence of rapidity divergences and UV divergences leads to the renormalization theorem for rapidity divergences (RTRD).

The relation between rapidity and UV divergences and RTRD have multiple consequences. The most important one is the factorization of rapidity divergences for soft factors. In the case of TMD factorization this statement is known, and thus, RTRD brings a little new, apart of some formality. However, it is novel for the double-parton scattering and MPS. The relation between different kind of divergences allows to relate the corresponding anomalous dimensions. In our case, it gives the correspondence between the soft anomalous dimension (SAD) and the rapidity anomalous dimension (RAD), which has been discovered in \cite{Vladimirov:2016dll}. The RTRD formulated in this article has a number of limitations. In particular, it is formulated for the soft factors that could be presented as a single T-ordered operator. Such soft factors arise in the processes with DY kinematics or annihilation kinematics. The status of factorization for processes with timelike separation is not clear. However, we show that the SIDIS TMD soft factor is equal to DY TMD soft factor, which was expected for a long time. 

The structure of the paper is following. In sec.\ref{sec:notation} we collect all necessary notation. In sec.\ref{sec:MPS_SF} we present the MPS soft factor, which is the main object of discussion. In particular, its all-order color structure and the explicit expression up to the three-loop order are given in sec.\ref{sec:color}. The derivation of this result is given in the appendix \ref{app:color}. The factorization of rapidity divergences at the fixed order (two-loop) is presented in sec.\ref{sec:N=2} and sec.\ref{sec:N=4}.  In sec.\ref{sec:rap_div} we explore the origin of rapidity divergences on the level of Feynman diagrams. We start from the classification of divergences in the one-loop example in the position space in sec.\ref{sec:1loop}. In sec.\ref{sec:geom_rap_div} we discuss a general case and associate the rapidity divergent parts with a particular spatial configuration. Namely, we show that the gluons radiated to/by the transverse (to a given lightlike direction) plane positioned at the infinity, produce rapidity divergences. In sec.\ref{sec:2loop} we illustrate the general statement by two-loop examples and present the graphical counting rules for rapidity divergences, which appears to be topologically similar to counting rules for UV divergences.  Section \ref{sec:RTRD_gen} is devoted to the formulation and the proof of RTRD. In particular, in sec.\ref{sec:Cnn} we introduce the transformation $C_{n\bar n}$ which distinguishes the rapidity divergences,  and in sec.\ref{sec:RTRD} we prove the theorem in a conformal theory and QCD. In sec.\ref{sec:consiquences}, some consequences, and applications of the theorem are presented. We discuss the definition of multi-parton distributions (which include the TMD distributions and double-parton distributions as particular cases) in sec.\ref{sec:MPS_fac}. The universality of TMD soft factor for DY and SIDIS process is proven in sec. \ref{sec:TMD_UNIVERSAL}. Finally, we discuss the correspondence between the soft anomalous dimension and the rapidity anomalous dimension and derive the three-loop rapidity anomalous dimension for TMD and MPS cases in sec.\ref{sec:corespondance}. Some additional materials are collected in the set of appendices.

\section{Notation and definitions}
\label{sec:notation}

In the most part of the paper, we discuss the kinematics with two selected lightlike directions. Conventionally, we denote these directions as $n$ and $\bar n$, with $n^2=\bar n^2=0,\qquad (n\cdot \bar n)=1$. The decomposition of a vector over light-cone components is defined as
\begin{eqnarray}
x^\mu=\bar n^\mu x^++ n^\mu x^-+x_\perp^\mu.
\end{eqnarray}
Consequently, the components of the vector $x$ are
$$
x^+=(n \cdot x),\qquad x^-=(\bar n \cdot x),\qquad (n \cdot x_\perp)=(\bar n\cdot x_\perp)=0,
$$
and the scalar product is
\begin{eqnarray}
(x\cdot y)=x^+ y^-+x^-y^++(x_\perp\cdot y_\perp),
\end{eqnarray}
i.e. subscript $\perp$ denotes the transverse part in the Minkowski space ($x_\perp^2<0$).

Throughout the text, we use the color matrix notation, see e.g.\cite{Catani:1996vz,Beneke:2009rj,Gardi:2009qi}. Namely, we use the bold font for the color-matrices, and multi-matrices, i.e. for objects with two sets of color indices. The color vectors, i.e. the objects with one set of color indices are written in a usual font. The convolution between such objects is denoted by $\times$-symbol, e.g. $A_{a_1 a_2}B^{a_1a_2,b_1b_2}=A\times \mathbf{B}$. The generators of the color gauge group are denoted by $\mathbf{T}_i^A$, where $i$ labels the gauge-group representation. If some representation sub-space is not specified, this part of a matrix is unity.

The main objects of the discussion are soft factors. By a soft factor we widely understand a vacuum matrix element of any product of Wilson lines. The Wilson line from the point $x$ to the point $y$ reads
\begin{eqnarray}
\pmb{[y,x]}=P\exp\(ig \int_x^y dz^\mu A_\mu^A(z)\mathbf{T}^A\),
\end{eqnarray}
where the path of integration is the straight line from $x$ to $y$. Under the gauge transformation Wilson lines transforms as
\begin{eqnarray}
\pmb{[y,x]}\to \mathbf{U}(y)\pmb{[y,x]}\mathbf{U}^\dagger (x).
\end{eqnarray}
The group representation of the Wilson line is carried solely by the generator. For example, it implies that quark and anti-quark Wilson lines differ only by the color representation (fundamental and anti-fundamental), but not by the path.

A typical soft factor that arises in the factorization theorems, is build of half-infinite Wilson lines, which are specified by the direction and the initial point. The half-infinite Wilson line that is rooted at the position $x$ and points in the direction $v$ is denoted as
\begin{eqnarray}
\mathbf{\Phi}_v(x)=\pmb{[v\infty+x,x]}=P\exp\(ig \int_0^\infty d\sigma v^\mu A_\mu^A(v\sigma+x)\mathbf{T}^A\).
\end{eqnarray}
The half-infinite Wilson line pointing in the opposite direction is
\begin{eqnarray}
\mathbf{\Phi}_{-v}(x)=\pmb{[-v\infty+x,x]}=P\exp\(ig \int_0^{-\infty} d\sigma v^\mu A_\mu^A(v\sigma+x)\mathbf{T}^A\).
\end{eqnarray}

In the most part of the article, the discussion is not restricted to any rapidity regularization. However, for the demonstrations of particular expressions we use the $\delta$-regularization. The synopsis of $\delta$-regularization and some of its properties are given in appendix \ref{app:delta-reg}.

\section{MPS soft factor}
\label{sec:MPS_SF}

The starting and the main object of our analysis is the soft factor of multi-Drell-Yan (multi-DY) process. Such a soft factor would appear in the description of hadron-hadron collision with the inclusive production of multiple heavy electro-weak bosons, e.g. $h_1+h_2\to Z_1+...+Z_N+X$ with the momenta of $Z$-bosons $Q_i\gg \Lambda_{QCD}$. The factorization theorem for multi-DY process contains many terms and various kinds of contributions. In particular, we are interested in the contribution which corresponds to the so-called multi-parton-scattering (MPS) subprocess. The MPS is characterized by the vector boson production by uncorrelated pairs of partons. The detailed discussion on these processes and possibilities to study them practically can be found in refs.\cite{Diehl:2011yj,Manohar:2012jr,Diehl:2015bca}. For our discussion, the multi-DY process is interesting as a generalization of the DY TMD factorization. It preserves the general structure of the factorization theorem and suffers from the same problem, namely the mix of rapidity divergences. The factorization of the MPS contribution of factorization theorem is discussed in sec.\ref{sec:MPS_fac}.

\begin{figure}[t]
\centering
\includegraphics[width=0.5\textwidth]{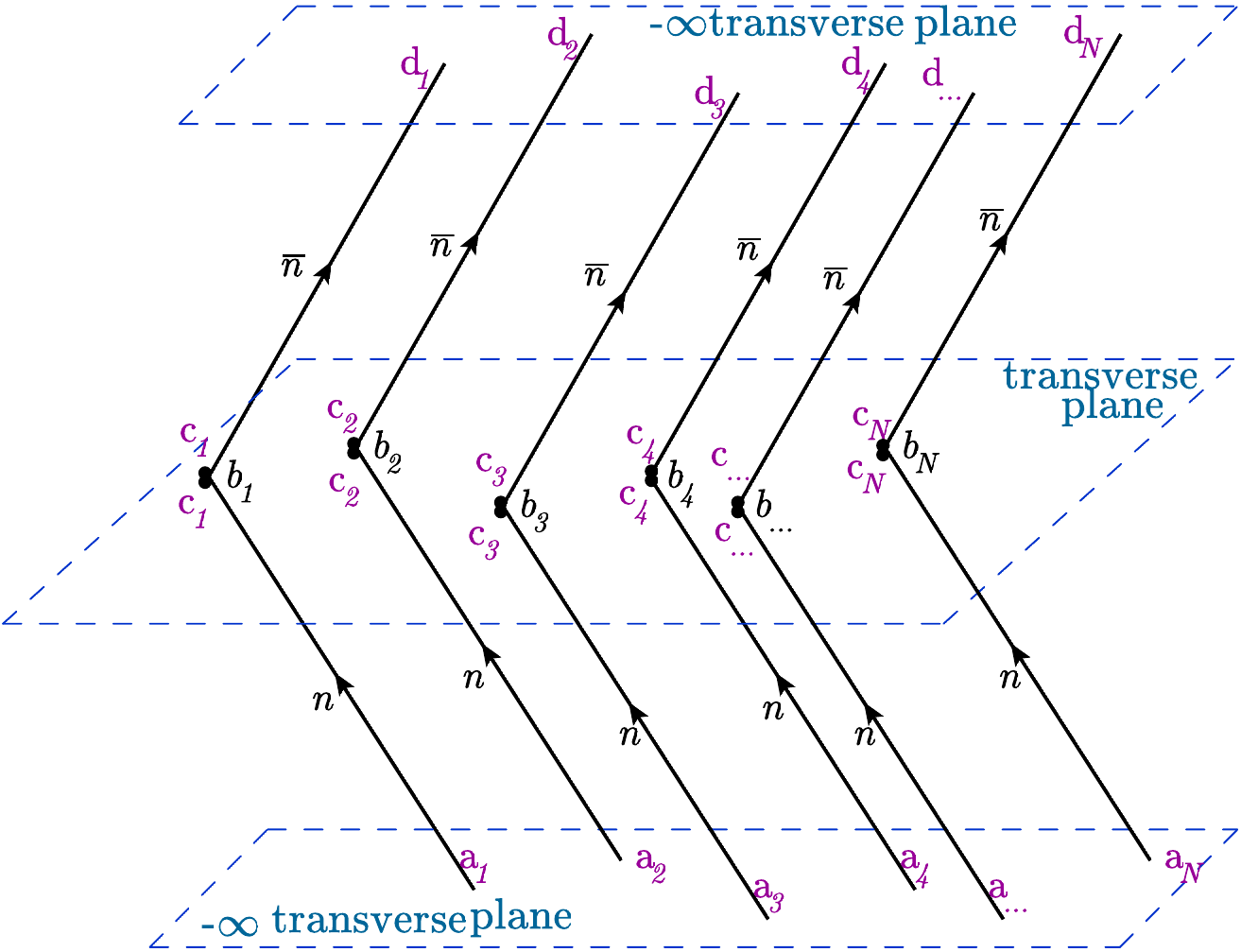}
\includegraphics[width=0.41\textwidth]{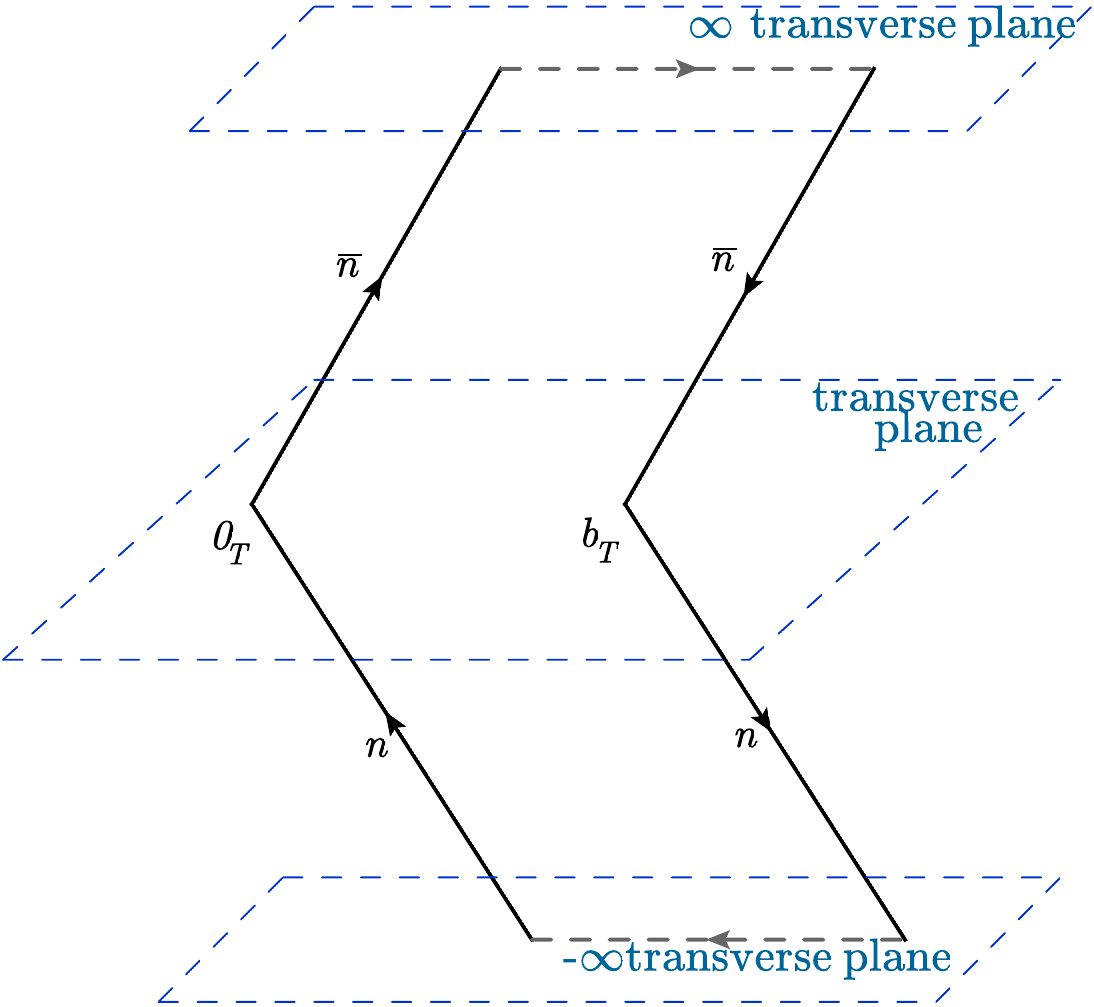}
\caption{\label{fig:MPS+TMD_SF} Visualization of the expression for the MPS soft factor (\ref{def:MPS_SF_openI}) (left) and the TMD soft factor (right). The lines with arrows represent the Wilson lines with the color-flow in the direction of the arrow. The purple letters show the color indices, the black letters show the transverse positions and directions.} 
\end{figure}

The soft factor for multi-DY process reads \cite{Diehl:2011yj} (in the following we call it MPS soft factor for shortness)
\begin{eqnarray}\label{SF:MPS_Tordered}
\mathbf{\Sigma}(\{b\})=\langle 0|\bar T\{[\mathbf{\Phi}_{-n}\mathbf{\Phi}^\dagger_{-\bar n}](b_N)\dots\}T\{\dots[\mathbf{\Phi}_{-n}\mathbf{\Phi}^\dagger_{-\bar n}](b_1)\}  |0\rangle,
\end{eqnarray}
where Wilson lines inside square brackets belong to the same color-representation and thus contracted by the internal index, and vectors $b$ have only transverse components, i.e. $b^+_i=b^-_i=0$. We stress that the MPS soft factor is a multi-matrix in the color space. To clarify the notation we write this expression with all color indices explicit
\begin{eqnarray}\label{def:MPS_SF_openI}
\Sigma^{\{a_N...a_1\},\{d_N...d_1\}}(\{b\})=\langle 0|\bar T\{[\Phi^{a_Nc_N}_{-n}\Phi^{\dagger c_N d_N}_{-\bar n}](b_N)\dots\}T\{\dots[\Phi^{a_1c_1}_{-n}\Phi^{\dagger c_1d_1}_{-\bar n}](b_1)\}  |0\rangle.
\end{eqnarray}
The MPS soft factor can be visualized as a set of lightlike cusps located at the transverse plane, as it is shown in fig.\ref{fig:MPS+TMD_SF}.

Only the color-singlet components of the soft factor matrix contribute to the factorization theorem. Generally, one can build the vector $C_K$ that selects the $K$'th singlet component. The complete set of vectors $C_K$ can be normalized and orthogonalized: $C^T_M\times C_N=\delta_{MN}$. The physically relevant part of the MPS soft factor reads
\begin{eqnarray}
\Sigma_{MN}(\{b\})=C^T_M\times \mathbf{\Sigma}(\{b\})\times C_N.
\end{eqnarray}
Only these components of the MPS soft factor are gauge-invariant, and IR-finite. Within the color-matrix notation the singlet components can be selected out by requiring (for the origin of this equation see e.g. \cite{Catani:1996vz,Beneke:2009rj})
\begin{eqnarray}\label{def:colorless}
\sum_{i=1}^N\mathbf{T}^A_i=0.
\end{eqnarray}
In the following, we consider only physical components, which pick out by the relation (\ref{def:colorless}).

To our best knowledge the MPS soft factor has been never considered in the literature (some discussion can be found in ref.\cite{Diehl:2011yj}). In the following subsections, we discuss some important cases and properties of MPS soft factor, including the all-order color structure and the explicit NNLO expression, which are presented here for the first time.

\subsection{T-ordering}

The soft factors for the DY-like kinematics have Wilson lines settled on the past-light-cone. It has an important consequence, which makes possible the following analysis. Namely, all Wilson lines within the soft factor operator can be set under the single T-ordering. I.e. the expression (\ref{SF:MPS_Tordered}) can be written as
\begin{eqnarray}\label{SF:MPS}
\mathbf{\Sigma}(\{b\})=\langle 0|T\{[\mathbf{\Phi}_{-n}\mathbf{\Phi}^\dagger_{-\bar n}](b_N)\dots[\mathbf{\Phi}_{-n}\mathbf{\Phi}^\dagger_{-\bar n}](b_1)\}  |0\rangle.
\end{eqnarray}
It can be demonstrated as follows: (i) The distances between points of any two Wilson lines are spacelike. And hence, the T- and anti-T-orderings can be ignored due to the causality condition. (ii) The path ordering of a Wilson line overrides\footnote{In the case of Wilson lines with timelike directions the path ordering and time ordering contradict each other. It can result to the extra non-physical singularities, in the self-interacting diagrams, see e.g. discussion in \cite{Collins:2011zzd}. For lightlike Wilson lines, which are discussed here, there is no such problem.} the anti-T ordering. Therefore, all Wilson lines can be T-ordered. (iii) Finally, using the causality we collect all Wilson lines under the single T-ordering.

The overall T-ordering of the operator is important for future discussion. It is not the general property for soft factors. For example, in the SIDIS kinematics, soft factors are built from $[\mathbf{\Phi}_{-n}\mathbf{\Phi}^\dagger_{\bar n}]$-cusps. In this case, not all distances are spacelike. Hence, the T-ordering cannot be eliminated. Another important example is the $e^+e^-$-annihilation, where soft factors are composed from $[\mathbf{\Phi}_{n}\mathbf{\Phi}^\dagger_{\bar n}]$-cusps, and can be presented as a single T-product.

\subsection{Particular case $N=2$: the TMD soft factor}
\label{sec:N=2}

The MPS soft factor at $N=2$ reduces to the TMD soft factor for the DY process, see e.g.\cite{Bauer:2001yt,Becher:2010tm,GarciaEchevarria:2011rb,Collins:2011zzd,Echevarria:2015byo}. In $N=2$ case, the color-neutrality condition (\ref{def:colorless}) relates the generators of the first and the second Wilson lines as $\mathbf{T}_1^A=-\mathbf{T}_2^A$. The matrix $\mathbf{\Sigma}$ has only single colorless entry $\sim\delta^{a_1a_2}/\text{dim}_1=I_{\mathbf{1}}$. Projecting the singlet contribution we obtain
\begin{eqnarray}\label{SF:TMD}
\Sigma_{\text{TMD}}(b)=I_{\mathbf{1}}\times \mathbf{\Sigma}_{N=2}(b)\times I_{\mathbf{1}} = \frac{1}{\text{dim}_1}
\langle 0|\bar T\{\Phi^{dc_2}_{-n}(b)\Phi^{\dagger c_2a}_{-\bar n}(b)\}T\{\Phi^{ac_1}_{-n}(0)\Phi^{\dagger c_1d}_{-\bar n}(0)\}  |0\rangle,
\end{eqnarray}
where one of the vectors $b$ is eliminated by the translation invariance. To derive this relation we have used the relation
\begin{eqnarray}
\mathbf{\Phi}_v(x)[-\mathbf{T}]=\mathbf{\Phi}^*_v(x)=\(\mathbf{\Phi}^\dagger_v(x)\)^T.
\end{eqnarray}
The visualization of the expression (\ref{SF:TMD}) is given in fig.\ref{fig:MPS+TMD_SF}.

The TMD soft factor is a Wilson loop. Therefore, the non-Abelian exponentiation theorem \cite{Gatheral:1983cz,Frenkel:1984pz} can be applied, and the soft factor takes the form
\begin{eqnarray}\label{SF:TMD->sigma}
\Sigma_{\text{TMD}}(b)=\exp\(C_1a_s \sigma(b)\),
\end{eqnarray}
where $C_1$ is the eigenvalue of the quadratic Casimir for the representation $1$, and $\sigma$ is given by the sum of the web-diagrams. The LO expression for $\sigma$ in the $\delta$-regularization reads
\begin{eqnarray}\label{SF:1loop}
\sigma^{[0]}=-4\Gamma(-\epsilon)(\mu^2 B)^\epsilon(L_\delta-\psi(-\epsilon)-\gamma_E),
\end{eqnarray}
where $a_s=g^2/(4\pi)^2$,  $B=b^2/4 e^{-2\gamma_E},$ and $ L_\delta=\ln(\delta^+\delta^- B)$. The parameters $\delta^+$ and $\delta^-$ regularize the rapidity divergences which arise due to the interaction with Wilson lines $\Phi_{-n}$ and $\Phi_{-\bar n}$ correspondingly. Obviously, the rapidity divergences belonging to different sectors can be split into separate functions by presentation of $\ln(\delta^+\delta^-)$ as $\ln\delta^++\ln\delta^-$. 

The explicit calculations performed in different regularizations \cite{Echevarria:2015byo,Luebbert:2016itl,Li:2016axz}, demonstrate that the linearity of the TMD soft factor in $L_\delta$ (or corresponding rapidity divergent function) holds at NLO as well. I.e.
\begin{eqnarray}\label{TMD:structure}
\sigma(b)=A(b,\epsilon)L_\delta+B(b,\epsilon),
\end{eqnarray}
where $A$ and $B$ are known up to $a_s^2$-order. As it is discussed in the introduction, the status of this formula at higher orders is not clear. However, the expression (\ref{TMD:structure}) is expected to hold at all orders of the perturbation theory. In particular, it holds at the leading order of large-$N_f$ expansion \cite{Scimemi:2016ffw}. Using the representation (\ref{TMD:structure}), the TMD soft factor can be written in the form
\begin{eqnarray}\label{TMD:FAC}
\Sigma_{\text{TMD}}(b,\delta^+,\delta^-)=\sqrt{\Sigma_{\text{TMD}}(b,\delta^+,\delta^+)}\sqrt{\Sigma_{\text{TMD}}(b,\delta^-,\delta^-)}.
\end{eqnarray}
This relation is the foundation for the TMD factorization and the definition of TMD distributions.

\subsection{Particular case $N=4$: DPS soft factor}
\label{sec:N=4}

\begin{figure}[t]
\centering
\includegraphics[width=0.45\textwidth]{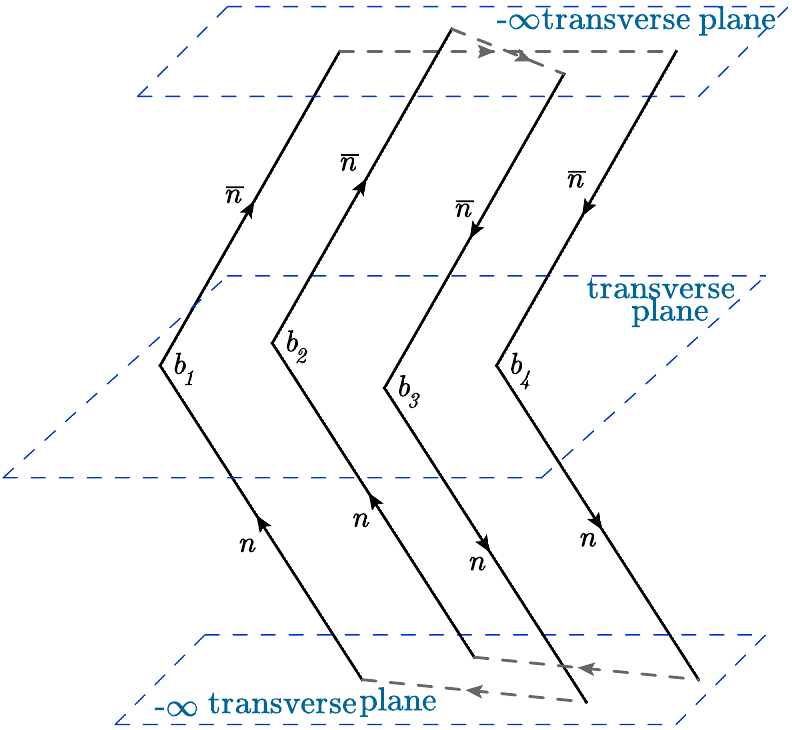}
\includegraphics[width=0.45\textwidth]{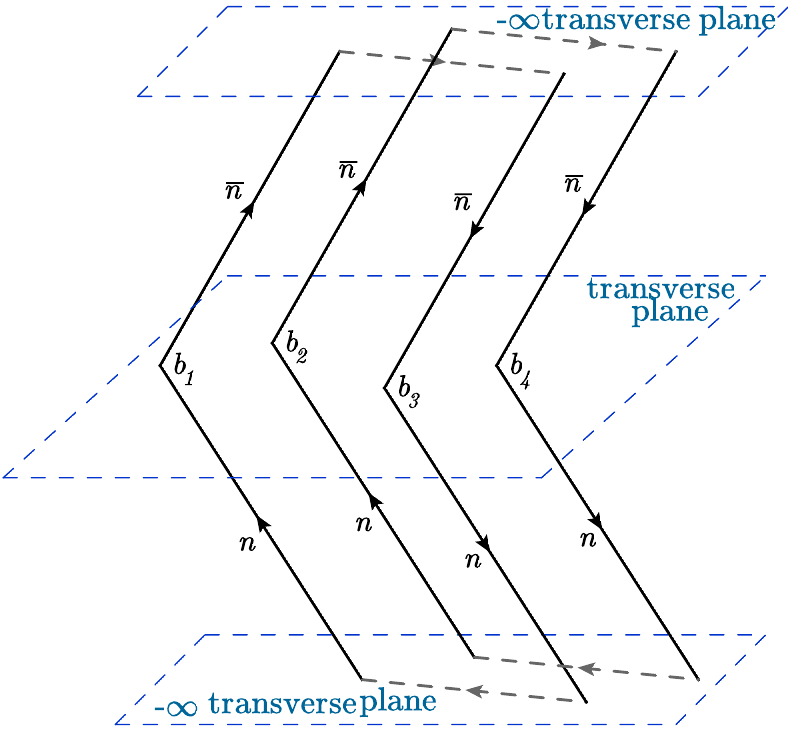}
\caption{\label{fig:DPD_SF} Visualization of $\Sigma^{[1]}_{\text{DPD}}$ (left), which has a topology of the single Wilson loop, and $\Sigma^{[2]}_{\text{DPD}}$(right), which has topology of the double Wilson loop.} 
\end{figure}

The $N=4$ case describes the double-patron-scattering (DPS) process. The details on the factorization theorems for this process can be found in \cite{Diehl:2011yj,Manohar:2012jr,Diehl:2015bca,Vladimirov:2016qkd}. There are many possible configurations for $N=4$ soft factors, which correspond to different parton content of the scattering subprocess. For the demonstration purpose, we present the simplest case of two quarks and two anti-quarks, which already gives six possibilities for the color flow. For definiteness, we present here the combination of \{fundamental, anti-fundamental, fundamental, anti-fundamental\} Wilson lines, that corresponds to \{quark,anti-quark,quark,anti-quark\} scattering or the contribution of double-parton distributions $F_{q\bar q}$. Such a composition is projected to the singlets by two vectors
\begin{eqnarray}\label{DPD:proj}
I_{\bf{1}}=\frac{\delta_{a_1a_4}\delta_{a_2a_3}}{N_c^2},\qquad I_{\bf{8}}=\frac{2t_{a_1a_4}^{A}t^A_{a_3a_2}}{N_c\sqrt{N_c^2-1}}.
\end{eqnarray}
Therefore, the DPD soft factor is the two-by-two matrix. Practically, it is convenient to present it in the following form \cite{Vladimirov:2016qkd}
\begin{eqnarray}\label{DPD:SFmatrix}
\Sigma_{\text{DPD}}(\{b\})&=&\(
\begin{array}{cc}
\Sigma^{\mathbf{11}}_{\text{DPD}}(\{b\})& \Sigma^{\mathbf{18}}_{\text{DPD}}(\{b\})
\\
\Sigma^{\mathbf{81}}_{\text{DPD}}(\{b\})& \Sigma^{\mathbf{88}}_{\text{DPD}}(\{b\})
\end{array}
\)
\\\nn
&=&
\frac{1}{N_c^2}\(\begin{array}{cc}
\Sigma^{[2]}_{\text{DPD}}(b_{1,3,4,2})& 
\frac{N_c\Sigma^{[1]}_{\text{DPD}}(b_{1,2,3,4})-\Sigma^{[2]}_{\text{DPD}}(b_{1,2,3,4})}{\sqrt{N_c^2-1}}
\\
\frac{N_c\Sigma^{[1]}_{\text{DPD}}(b_{1,2,3,4})-\Sigma^{[2]}_{\text{DPD}}(b_{1,2,3,4})}{\sqrt{N_c^2-1}} & 
\frac{N_c^2\Sigma^{[2]}_{\text{DPD}}(b_{1,4,3,2})-2N_c\Sigma^{[1]}_{\text{DPD}}(b_{1,2,3,4})+\Sigma^{[2]}_{\text{DPD}}(b_{1,2,3,4})}{N_c^2-1}
\end{array}\),
\end{eqnarray}
where arguments $b_{i,j,k,l}$ are short notation for $(b_i,b_j,b_k,b_l)$, and $\Sigma^{\mathbf{ij}}_{\text{DPD}}(\{b\})=I_{\mathbf{i}}\times\mathbf{\Sigma}_{N=2}(\{b\})\times I_{\mathbf{i}}$. The soft factors $\Sigma^{[1]}_{\text{DPD}}$ and $\Sigma^{[2]}_{\text{DPD}}$ are soft factors with Wilson lines connected into single and double color loop, see fig.\ref{fig:DPD_SF}.

The explicit calculation of the DPS soft factor at NNLO has been made in \cite{Vladimirov:2016qkd}. It has been shown that DPS soft factor has a number of peculiarities. The most important one is the exact cancellation of the three-Wilson line interactions. Due to this cancellation the NNLO soft factor can be expresed via the TMD soft factor only. The result is not entirely trivial. The single-loop and double-loop components are
\begin{eqnarray}\label{DPD:NNLO1}
&&\ln\Sigma^{[1]}_{\text{DPD}}(\{b\})=a_sC_F\(\sigma(b_{12})-\sigma(b_{13})+\sigma(b_{14})+\sigma(b_{23})-\sigma(b_{24})+\sigma(b_{34})\)
\\\nn&&\qquad\quad+a_s^2\frac{C_A C_F}{4}\(\sigma(b_{13})-\sigma(b_{14})-\sigma(b_{23})+\sigma(b_{24})\)
\(\sigma(b_{12})-\sigma(b_{13})-\sigma(b_{24})+\sigma(b_{34})\)+\mathcal{O}(a_s^3),
\\\label{DPD:NNLO2}
&& \ln\Sigma_{\text{DPD}}^{[2]}(\{b\})=a_sC_F\(\sigma(b_{14})+\sigma(b_{23})\)
\\\nn&&\qquad\quad-a_s^2\frac{C_F}{2}\(C_F-\frac{C_A}{2}\)\(\sigma(b_{12})-\sigma(b_{13})-\sigma(b_{24})+\sigma(b_{34})\)^2+\mathcal{O}(a_s^3),
\end{eqnarray}
where $\sigma(b)$ is the logarithm of the TMD soft factor (\ref{SF:TMD->sigma}), and $b_{ij}=b_i-b_j$. One can see that these components have double rapidity logarithms which do not cancel.

Combining the expression for components (\ref{DPD:NNLO1}-\ref{DPD:NNLO2}) into the matrix of the DPS soft factor (\ref{DPD:SFmatrix}) one obtains a complicated expression. However, this expression can be presented in the form
\begin{eqnarray}\label{DPD:FAC}
\Sigma_{\text{DPD}}(\{b\},\delta^+,\delta^-)=s^T(\{b\},\delta^+)s(\{b\},\delta^-),
\end{eqnarray}
where $s^T$ is the transposed matrix $s$. The expression for matrices $s$ is cumbersome (see \cite{Vladimirov:2016qkd}) but it has a general form
\begin{eqnarray}\label{DPD:small_s}
s(\{b\},\delta)=\exp\(\mathcal{A}(\{b\})L_{\delta}+\mathcal{B}(\{b\})\),
\end{eqnarray}
where the $2\times 2$ matrices $\mathcal{A}$ and $\mathcal{B}$ are composed from functions $A$ and $B$ defined in (\ref{TMD:structure}). 

The decomposition (\ref{DPD:FAC}) is the matrix generalization of the TMD decomposition formula (\ref{TMD:FAC}). It defines the finite double parton distribution (DPD) in the very same manner as the decomposition (\ref{TMD:FAC}) defines TMD distributions (see details in sec.\ref{sec:MPS_fac}). The main difference between (\ref{TMD:FAC}) and (\ref{DPD:FAC}) is the matrix structure, which leads to the matrix rapidity evolution equation for DPDs. In the next section, we demonstrate the generalization of this expression for the MPS soft factor. However, we also demonstrate that at the three-loop level the new types of terms appear, that do not reduce to the TMD soft factors. The analysis of these terms is difficult, and their rapidity divergences structure is unknown. 

\subsection{Color structure}
\label{sec:color}

The MPS soft factor has reach color structure. Its expression is greatly simplified in the color matrix notation. In the appendix \ref{app:color} we present the detailed evaluation of the MPS soft factor in the terms of the generating functions for web-diagrams \cite{Vladimirov:2015fea,Vladimirov:2014wga}. Such decomposition extracts the color structure explicitly, and reveals the common structure of diagrams. In this section, we present the final result of the decomposition. 

The first and the most important observation on the color structure of MPS soft factor follows from the rotation invariance. Performing the rotation that interchange $n\leftrightarrow \bar n$, we obtain
\begin{eqnarray}
\mathbf{\Sigma}(\{b\})=\mathbf{\Sigma}^\dagger(\{b\}).
\end{eqnarray}
For the generator of the color group this transformation acts as $\mathbf{T}_i\to -\mathbf{T}_i$. Therefore, the terms with the odd-number of color-generators vanish. This statement also holds for the exponentiated expression (see (\ref{app:odd=0}) and the discussion around). This observation describes the absence of the three-Wilson lines interaction terms in the DPS soft factor (\ref{DPD:NNLO1}-\ref{DPD:NNLO2}), which has been shown on the level of diagrams in \cite{Vladimirov:2016qkd}. The general all-order structure of MPS soft factor reads
\begin{eqnarray}\label{MPS:all_order_color}
\mathbf{\Sigma}(\{b\})=\exp\(\sum_{\substack{n=2\\n\in\text{even}}}^{\infty}a_s^{n/2}\sum_{i_1,...,i_n=1}^N\{\mathbf{T}^{A_1}_{i_1}...\mathbf{T}^{A_n}_{i_n}\}\sigma^{n;i_1...i_n}_{A_1...A_n}(\{b\})\),
\end{eqnarray}
where curly brackets denote the symmetrization over the color generators belonging to the same Wilson lines. The functions $\sigma^n\sim \mathcal{O}(a^0_s)$ obey the same symmetry pattern under permutation of labels $i$ and $A$ as the color structure. Note, that all matrices belonging to the same Wilson lines appear in the symmetric combinations. The anti-symmetric combinations are absent due to the algebra. The $n=0$ term is rewritten as $n=2$ contribution using the color-conservation. 

The number of independent color components grows rapidly with order. However, their number is finite at given $N$, thanks to color-algebra and color-conservation condition (\ref{def:colorless}). In particular, at the three-loop level there are three independent structures
\begin{eqnarray}\label{SF:color_dec}
\mathbf{\Sigma}&=&\exp\Bigg[-a_s \sum_{[i,j]}\mathbf{T}_i^A\mathbf{T}_j^A \sigma(b_{ij})
\\\nn&&
+a_s^3 \Big(\sum_{[i,j,k]}\mathbf{T}_i^{\{AB\}}\mathbf{T}_j^C\mathbf{T}_k^D if^{AC;BD}Y_{4}^{ijk}
+\sum_{[i,j,k,l]}\mathbf{T}_i^{A}\mathbf{T}_j^B\mathbf{T}_k^C\mathbf{T}_l^D if^{AC;BD}X_{4}^{ijkl}\Big)+\mathcal{O}(a_s^4)\Bigg],
\end{eqnarray}
where $if^{AC;BD}=if^{AC\alpha}if^{\alpha BD}$, with $f^{ABC}$ being the structure constant, $b_{ij}=b_i-b_j$, and $\mathbf{T}_i^{\{AB\}}=\{\mathbf{T}_i^A,\mathbf{T}_i^B\}/2$. Here, the summation runs from $1$ to $N$ for each summation label with no label equals to any other label, which is denoted by the square brackets. Functions $\sigma$, $Y_4$, and $X_4$ contain all orders of perturbation series starting from the LO. Their explicit form in the terms of generating function is given in section \ref{app:results}. The expression (\ref{SF:color_dec}) is simpler then the expressions (\ref{DPD:SFmatrix}-\ref{DPD:NNLO2}), to which they turn after application of projectors (\ref{DPD:proj}).

The color-dipole term in the decomposition (\ref{SF:color_dec}) is proportional to the TMD soft factor, which can be checked by setting $N=2$. Assuming the linearity of $\sigma$ in $\ln(\delta^+\delta^-)$ at all-orders of the perturbation theory and also the linearity of $X_4$ and $Y_4$ in $\ln(\delta^+\delta^-)$ we can present this expression in the factorized form (\ref{DPD:FAC}) as well (up to terms $\sim a_s^4$). 

\section{Divergences of soft factors}
\label{sec:rap_div}

The soft factors with lightlike Wilson lines are utterly singular objects. Diagram-by-diagram there are UV-, IR-, and rapidity divergences. To define the soft factor completely, a sufficient set of regulators should be introduced. Typically, it includes the dimensional regularization for UV- and IR-divergences, and an extra regulator for the rapidity divergences. In this section we discuss the diagrammatic origin of the rapidity divergences, and show their similarity to the UV divergences.

\subsection{Divergences of soft factor at one loop}
\label{sec:1loop}

To begin with let us consider the LO contribution to the interaction of Wilson lines. At LO there could be many diagrams (depending on the structure of the soft factor, the gauge conditions and calculation technique). However, there is a single loop-integral that appears at this order. This integral describes the single-gluon exchange between Wilson lines $\Phi_{v_i}(b_i)$ and $\Phi_{v_j}(b_j)$ (for the demonstration purpose we keep vectors $v$ and $b$ unrestricted). In the coordinate representation, it reads
\begin{eqnarray}\label{1loop:1}
I_{ij}&=&
a_s 2^{2-2\epsilon}\Gamma(1-\epsilon)\int_0^\infty d\sigma_1 d\sigma_2\frac{(v_i\cdot v_j)}{(-(v_i\sigma_1+b_i-v_j\sigma_2-b_j)^2+i0)^{1-\epsilon}},
\end{eqnarray}
where $a_s=g^2/(4\pi)^{2-\epsilon}$. We use the dimensional regularization with $d=4-2\epsilon$, and do not specify the rapidity divergence regulator.

\begin{figure}[t]
\centering
\includegraphics[width=0.50\textwidth]{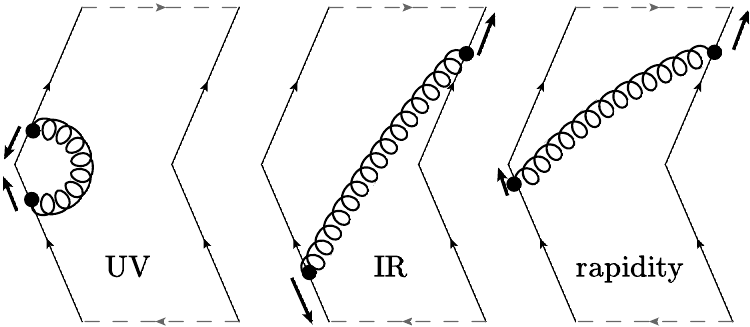}
\caption{\label{fig:divergences} Divergent configurations of the TMD soft factor at one-loop. The arrows indicate the direction in which the position of the particle should be limited.}
\end{figure}

In the expression (\ref{1loop:1}) the variables $\sigma$ represent the distances of gluon radiation/absorption along Wilson line. It is convenient to change the variables as $\sigma_1=\alpha L$ and $\sigma_2=\alpha^{-1} L$. In these terms, the variable $L$ represents the general ``size'' of the loop, and the variable $\alpha$ represents the $n/\bar n$-asymmetry of the gluon positioning. The integral (\ref{1loop:1}) takes the form
\begin{eqnarray}
&&I_{ij}=
a_s 2^{2-2\epsilon}\Gamma(1-\epsilon)\int_0^\infty dL \int_0^\infty d\alpha\frac{2L}{\alpha}
\\\nn &&\quad\frac{(v_i\cdot v_j)}{(-(v_i^2\alpha^2-2(v_i\cdot v_j)+v_j^2\alpha^{-2})L^2-2(v_i \alpha -v_j \alpha^{-1})\cdot (b_i-b_j)L-(b_i-b_j)^2+i0)^{1-\epsilon}}.
\end{eqnarray}
Let us sort the singularities of this integral, and depict the corresponding space configurations. Starting from the obvious:
\begin{itemize}

\item[]\textit{UV divergence}. In the case $b_i=b_j$, there is UV singularity at $L\to 0$. The integral behaves as $I\sim L^{-1+2\epsilon}$, and is regularized by $\epsilon>0$. The UV divergence is a subject of the  usual renormalization procedure.

\item[]\textit{IR divergence}. For any configuration one has IR singularity at $L\to \infty$. The integral behaves as $I\sim L^{-1+2\epsilon}$, and is regularized by $\epsilon<0$. For color singlet configurations the IR singularities cancel in the sum of diagrams. At LO the cancellation of IR-singularities is evident. Indeed, in the limit $L\to\infty$ vectors $b$ drop from the integral, and thus all IR-divergent integrals are equivalent. The proof of the cancellation at arbitrary perturbative  order is given in the appendix \ref{sec:delta-structure}.

\item[]\textit{Rapidity divergence}. In the special case, $v_i^2=v_j^2=0$ and $v_i\cdot (b_i-b_j)=v_j\cdot (b_i-b_j)=0$ the integral over $\alpha$ decouples from the integral over $L$, 
\begin{eqnarray}\label{rapdiv:1-loop}
I_{ij}&=&a_s 2^{2-2\epsilon}\Gamma(1-\epsilon) \int_0^\infty d L
\frac{2L(v_i\cdot v_j)}{(2(v_i\cdot v_j)L^2-(b_i-b_j)^2+i0)^{1-\epsilon}}\int_0^\infty \frac{d\alpha}{\alpha}.
\end{eqnarray}
The integral over $\alpha$ is logarithmically divergent at both limits $\alpha \to0$ and $\alpha\to \infty$. Such singularity is called the rapidity divergence.
\end{itemize}
The visual representation of the divergent configurations for the case of TMD soft factor is shown in fig.\ref{fig:divergences}.

The rapidity divergences are present even if a single vector $v_i$ is lightlike and orthogonal to the rooting plane, i.e. $v_i^2=0$ and $v_i\cdot (b_i-b_j)=0$ (and the second vector $v_j$ is arbitrary). In this case, the integral is regular at $\alpha\to0 $, but divergent at $\alpha\to \infty$. Moreover, the coefficient of this divergence is just the same as in (\ref{rapdiv:1-loop}).

\subsection{Spatial structure of rapidity divergences}
\label{sec:geom_rap_div}

In the one-loop example, the rapidity divergence arises from the integration over the half-infinite path of lightlike Wilson lines. Let us demonstrate that it is a general feature, and the rapidity divergence can arise for each coupling of the gluon to $\mathbf{\Phi}$. Note, that it is difficult to present the strict definition of rapidity divergences, because they are related to a particular component of gluon fields, and therefore, depend on the gauge fixation condition. In the following, we use the Feynman gauge for simplicity. 

A general diagram with a single gluon radiated by $\mathbf{\Phi}_{v}(b)$ ($v^2=0$) has the following form in the coordinate representation
\begin{eqnarray}
I^{[1]}&=&\int_0^\infty d\sigma \int d^d y \frac{1}{(-(v\sigma+b-y)^2+i0)^{p}}F(y),
\end{eqnarray}
where $p$ is the power of propagator that connects the Wilson line with the rest of the diagram which is denoted by $F(y)$. The function $F(y)$ can have its own divergences which are not interesting in the current context. For a given lightlike vector $v$, we introduce the decomposition
\begin{eqnarray}\label{rapdiv:y_decomp}
y^\mu=v^\mu y_s+s^\mu y_v+y_\perp^\mu,
\end{eqnarray}
where $(v\cdot y_\perp)=0$. Without loss of generality we can set $(v\cdot s)=1$. The components $y_v$, $y_s$, and $y_\perp$ are independent, and $d^dy=dy_vdy_sd^{d-2}y_\perp$. Rescaling variables 
$$y_v\to \frac{y_v}{\sigma}+(v\cdot b),$$ 
we obtain
\begin{eqnarray}\label{rapdiv:ex1}
I^{[1]}&=&\int_0^\infty \frac{d\sigma}{\sigma}\int d y_sdy_v d^{d-2}y_\perp
\\\nn &&\frac{F(y_s,y_v/\sigma+(v\cdot b),y_\perp)}{\[2 y_v-
\(b-y_\perp-(v\cdot b)\)^2-\frac{2y_v}{\sigma}\(y_s+(s\cdot y_\perp)+s^2(v\cdot b)-(s\cdot b)\)-\frac{y_v^2 s^2}{\sigma^2}+i0\]^{p}},
\end{eqnarray}
Here, the rapidity divergence appears in the limit $\sigma\to \infty$, where the expression (\ref{rapdiv:ex1}) takes the form
\begin{eqnarray}
I^{[1]}_{\text{rap.div.}}&=&\int^\infty \frac{d\sigma}{\sigma}\int d y_sdy_v d^{d-2}y_\perp
\frac{F(y_s,(v\cdot b),y_\perp)}{\[2 y_v-
\(b-y_\perp-(v\cdot b)\)^2+i0\]^{p}}.
\end{eqnarray}
Note, that the divergent factor decouples from the rest of the diagram. Such configuration corresponds to the radiation of a gluon from the \textit{transverse} to $v^\mu$ plane to the far end of the Wilson line $\mathbf{\Phi}_v$.

If there are several gluons coupled to the Wilson line $\mathbf{\Phi}_v$ we can perform the rescaling for each coupled coordinate $y_i$ and obtain the rapidity divergent configurations. The power of rapidity divergence is at most equal to the number of gluons coupled to $\mathbf{\Phi}$'s. We  should also take into account that the coupling of gluons to a Wilson line is ordered, e.g. for three coupled gluons we have the integral $\int^\infty d\sigma_1\int^{\sigma_1} d\sigma_2\int^{\sigma_2} d\sigma_3$. Thus the limits $\sigma_i\to \infty$ should be taken in the same order, which however could be impossible due to the internal structure of the function $F$. In particular, such situation appears if the coordinate $y$ coupled to another Wilson lines (see the examples given in the next section).

To summarize the geometry of rapidity divergent configuration, we introduce special notation. Let us denote by $(v)_\perp^y$ the two-dimensional (or $(d-2)$-dimensional) plane which is transverse to $v$ and intersects the axis $v$ at the coordinate $y$. The rapidity divergences arise in the configuration with the gluon is radiated within the plane $(v)_\perp^y$ and absorbed within the plane $\lim_{\sigma\to \infty}(v)_\perp^\sigma=(v)_\perp^\infty$. In other words, the rapidity divergences associated with the gluons that are localized in the space between $(v)_\perp^y$ and $(v)_\perp^\infty$. Since the particular value of $y$ has no sense, we can relate rapidity divergences to the plane $(v)_\perp^\infty$ for simplicity. 

If there are several Wilson lines pointing in the same lightlike direction, which is the typical situation, then they share $(v)_\perp^\infty$. The rapidity divergences of this configuration are shared. They can be regularized by a single regularization parameter, and should not be distinguished. If there are several sets of Wilson lines with directions $v_i$, then there are also several planes $(v_i)_\perp^\infty$. If these planes do not intersect then the associated rapidity divergences do not overlap. In this case, they can be regularized separately (and as we show later separately renormalized). If the planes $(v_i)_\perp^\infty$ intersect then the rapidity divergences overlap and could not be separated. Fortunately, soft factors with such geometry do not appear practically. Important to note, that the definition of the transverse plane is not unique, since the vector $s^\mu$ which specifies the plane, has not unique definition.

\subsection{Two loop examples and counting of rapidity divergences}
\label{sec:2loop}

In this section, we give some two-loop examples of rapidity divergences, and specify their counting.

\begin{figure}[t]
\centering
\includegraphics[width=0.83\textwidth]{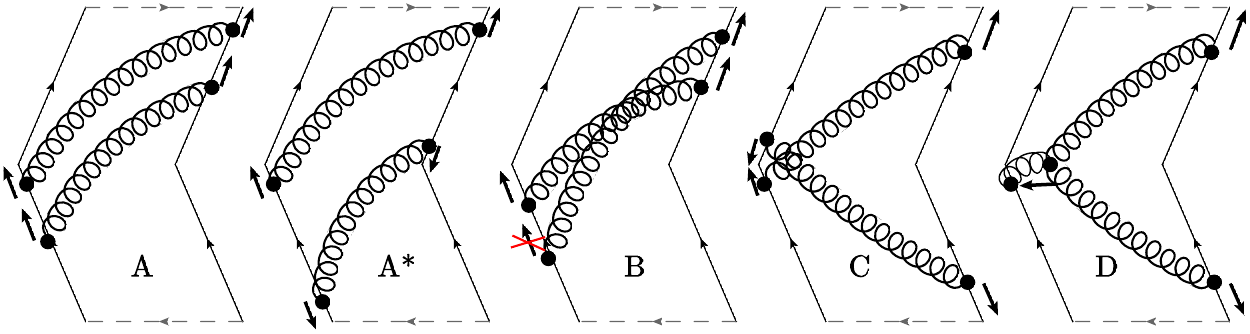}
\caption{\label{fig:divergences2} Examples of two-loop diagrams studied in the text. The diagrams $A$, $C$ and $D$ has the second power of rapidity divergence which appears if the positions of vertices are limited according to arrows. The diagram $B$ has the first power of rapidity divergence, since the positions of vertices cannot be limited according to arrows successively. The other combinations of divergent limits possible.}
\end{figure} 

As it was shown in the previous section, the overall power of the rapidity divergence in a diagram could not exceed the number of gluons attached to $\mathbf{\Phi}$'s. The maximum power of the divergences is achieved if all limits $\sigma_i\to \infty$, can be taken successively and decoupled from each other. This however is limited by the structure of the rest of the diagram. For example, it cannot be done if the divergent gluon is coupled to another Wilson line.

Let us give an example of similar diagrams which produce different power of rapidity divergences due to ordering of limits. These diagrams are shown in fig.\ref{fig:divergences2} A and B, and given by similar expressions (we omit the prefactors of loop-integrals for brevity)
\begin{eqnarray}
I_A&=&\int_0^\infty d\sigma_1\int_0^{\sigma_1} d\sigma_2 \int_0^\infty d\tau_1\int_0^{\tau_1} d\tau_2 \frac{1}{(2\sigma_1 \tau_2-b_{12}^2)^{1-\epsilon}
(2\sigma_2 \tau_1-b_{12}^2)^{1-\epsilon}},
\\\nn
I_B&=&\int_0^\infty d\sigma_1\int_0^{\sigma_1} d\sigma_2 \int_0^\infty d\tau_1\int_0^{\tau_1} d\tau_2 \frac{1}{(2\sigma_1 \tau_1-b_{12}^2)^{1-\epsilon}
(2\sigma_2 \tau_2-b_{12}^2)^{1-\epsilon}},
\end{eqnarray}
where $b_{12}$ is the transverse distance between lines. To extract the divergences associated with $\sigma\to \infty$, we rescale $\tau$ and obtain
\begin{eqnarray}
I_A&=&\int_0^\infty d\tau'_1\int_0^{\frac{\sigma_1}{\sigma_2}\tau'_1} d\tau'_2 \frac{1}{(2\tau'_2-b_{12}^2)^{1-\epsilon}
(2\tau'_1-b_{12}^2)^{1-\epsilon}}\int_0^\infty \frac{d\sigma_1}{\sigma_1}\int_0^{\sigma_1} \frac{d\sigma_2}{\sigma_2},
\\\nn
I_B&=& \int_0^\infty d\tau'_1\int_0^{\frac{\sigma_2}{\sigma_1}\tau'_1} d\tau'_2 \frac{1}{(2\tau'_1-b_{12}^2)^{1-\epsilon}
(2 \tau'_2-b_{12}^2)^{1-\epsilon}}\int_0^\infty \frac{d\sigma_1}{\sigma_1}\int_0^{\sigma_1} \frac{d\sigma_2}{\sigma_2}.
\end{eqnarray}
In the integral $I_A$ the limit $\sigma_1\to \infty$ decouples from the limit $\sigma_2\to \infty$ and we obtain the second power of rapidity divergence, as $(\int^\infty d\sigma/\sigma)^2$. In the integral $I_B$, the limit $\sigma_1\to \infty$ neglects the expression, and thus there is only single rapidity divergence which appears if both $\sigma$'s are sent to infinity simultaneously. The visual representation of the rapidity divergent configurations is given in  fig.\ref{fig:divergences2}A and B.

The diagram $A$ has the overlap of rapidity divergences associated with different directions. It appears in the limit $\sigma_1\to \infty $ and $\tau_2\to \infty$, which can be taken independently. It gives the rapidity divergences in both direction, $\big(\int^\infty d\sigma/\sigma\big) \big(\int^\infty d\tau/\tau\big)$. The corresponded geometrical configuration is shown in fig.\ref{fig:divergences2} A$^*$. In the $\delta$-regularization these substructures of diagram combine together into the Lorentz invariant expression $\sim \ln^2\delta^+\delta^-$ (here, $\ln^2 \delta^+$ corresponds to the double divergence in the $n$-direction, $\ln(\delta^+)\ln(\delta^-)$ to the mixed divergences and so on.) 

The diagram $A$ does not contribute to the TMD soft factor. It is not a web diagram  and thus, it is eliminated by the exponentiation procedure. However, in the case of the TMD soft factor, there are two other diagram topologies that give the double rapidity divergences. These diagrams are shown in fig.\ref{fig:divergences2} C and D. The explicit expression for these diagrams can be found e.g. in \cite{Echevarria:2015byo}. These diagrams have the same leading rapidity divergent structure proportional to $B^{2\epsilon}\Gamma^2(-\epsilon)\ln^2(\delta^+\delta^-)$, in the $\delta$-regularization. These double divergences cancel in the soft factor due to the different sign of the color coefficients. Note, that the diagram $C$ is simply a square of one-loop diagrams. The diagram $D$ has a more complicated expression, which can be reduced to the product of one-loop integrals in the rapidity divergent limit. Let us mention, that to obtain the rapidity divergent configuration in the diagram $D$ one of the vertices on the Wilson lines should be sent to the origin, while another two to infinity. We do not present the derivation here and refer the reader to ref.\cite{Erdogan:2011yc} where a similar evaluation (with the only absence of vector $b$) is performed. 

The examples that are given here confirm the general conclusion made in the previous section: The rapidity divergences are associated with gluons localized at $(v)_\perp^\infty$. To count the maximum power of rapidity divergence for a given diagram, one should draw the diagram and move the end point of a gluon attached to $\mathbf{\Phi}$ towards infinity, while the opposite side of this gluon is to be moved toward the origin (here we expect the "two-dimensional" TMD-like configuration of Wilson lines). If a gluon (or a subgraph) can be moved to the rapidity divergent limit without affecting the rest of the diagram, it decouples. The number of the vertices sent to infinities corresponds to the power of rapidity divergence. It is straightforward to show that the absolute maximum power of rapidity divergence does not exceed the number of coupling to Wilson lines, or the number of loops, whatever is smaller. 

The graph-topological structure of rapidity divergences reminds the graph-topological structure of UV divergences. The only difference is that positions of gluon couplings for UV divergent subgraphs should be limited to the same point, while for rapidity divergent subgraphs they should be limited to separate transverse planes. As we discuss in the next section it is not accidental, but the result of the fundamental relation between rapidity and UV divergences. Since the rapidity divergences have the same structure of the sub-divergences, we expect that they can be iterated by the Ward identities in a similar manner as the cusp UV divergence (or UV divergence of multi-cusp for the case of the MPS soft factor). Here we again refer to the detailed calculation made in ref.\cite{Erdogan:2011yc}, which can be nearly one-to-one repeated for the TMD soft factor. In fact, we expect that the renormalization theorem for rapidity divergences presented in the following sections can be proved in much the same way as the UV renormalization of the Wilson line cusp \cite{Dotsenko:1979wb,Brandt:1981kf}, i.e. by solving the chain of Ward identities.

\section{Renormalization theorem for rapidity divergences}
\label{sec:RTRD_gen}
\subsection{Conformal transformations of soft factor}
\label{sec:Cnn}

The rapidity divergences in many aspects resemble the UV divergences. The main difference is that the rapidity divergences are associated with the localization of gluons at the distant transverse plane $(-n)_\perp^{\infty}$, while the UV divergences are associated with the localization at a point. Let us build the conformal transformation which relates the plane $(-n)_\perp^{\infty}$ to a point (for simplicity we take the origin). It can be made by the following chain of transformations: (i) translation by $\{\frac{\lambda-1}{2 a},0^-,0_\perp\}$, (ii) special conformal transformation along the light-cone direction $n$ with the vector $\{0^+,a,0_\perp\}$ (iii) translation by $\{-(2 a)^{-1},0^-,0_\perp\}$. The resulting transformation reads
\begin{eqnarray}
\mathcal{C}_{\bar n}: \{x^+,x^-,x_\perp\}\to\{\frac{-1}{2 a}\frac{1}{\lambda+2 a x^+},x^-+\frac{a x_\perp^2}{\lambda+2 a x^+},\frac{x_\perp}{\lambda+2 a x^+}\}.
\end{eqnarray}
In the same manner we can build the transformation that relates the $(-\bar n)_\perp^\infty$ to the origin,
\begin{eqnarray}
\mathcal{C}_{n}: \{x^+,x^-,x_\perp\}\to\{x^++\frac{\bar a x_\perp^2}{\bar \lambda+2 \bar a x^-},\frac{-1}{2 \bar a}\frac{1}{\bar \lambda+2 \bar a x^-},\frac{x_\perp}{\bar \lambda+2 \bar a x^-}\}.
\end{eqnarray}
The parameters $a$ and $\lambda$ are free real parameters.

The combined transformation
\begin{eqnarray}
C_{n\bar n}=\mathcal{C}_n\mathcal{C}_{\bar n}=\mathcal{C}_{\bar n}\mathcal{C}_{n},
\end{eqnarray}
has a number of useful properties. 
The main geometric elements of the soft factor transform as
\begin{eqnarray}\nn
&&C_{n\bar n}(-\bar n)_\perp^\infty = \{0^+,\frac{-1}{2\bar a \bar \lambda},0_\perp\},
\\
&&C_{n\bar n}(-\bar n)_\perp^0 = C_{n\bar n}(-\bar n)_\perp^0 = S,
\\\nn
&&C_{n\bar n}(-n)_\perp^\infty = \{\frac{-1}{2 a \lambda},0^-,0_\perp\},
\end{eqnarray}
where $S$ is the two-dimensional surface
$$ S(y)=\frac{1}{\lambda\bar \lambda-2a\bar a y_T^2}\{\frac{-\bar\lambda}{2 a},\frac{-\lambda}{2 \bar a},y_T\},$$
with $y_T$ being arbitrary two-dimensional (Euclidean) vector. 

One can see that the plane $S$ is made by the intersection of two light-cones that are set at points $\{0^+,\frac{-1}{2\bar a \bar \lambda},0_\perp\}$ and $\{\frac{-1}{2 a \lambda},0^-,0_\perp\}$. The light-cones intersect by upper and lower branches, which form two disconnected branches of the surface $S$, parametrized by a single vector $y_T$. The boundary of the branch is determined by the equation $\lambda\bar \lambda=2a\bar a y_T^2$. Depending on the values of parameters $a$ and $\lambda$ the transformation realizes various configurations.

\begin{figure}[t]
\centering
\includegraphics[width=0.80\textwidth]{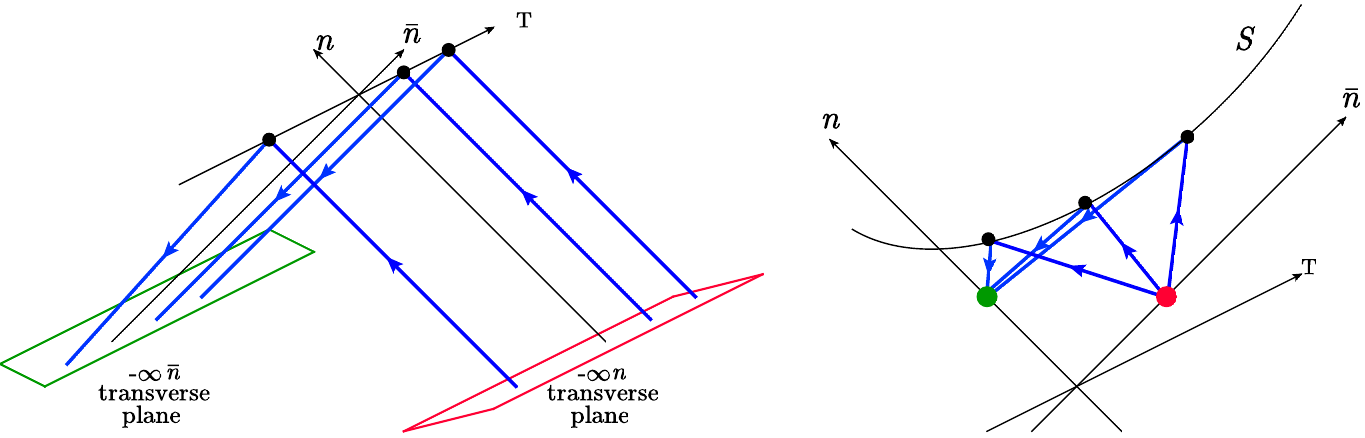}
\caption{\label{fig:Cnn} The shape of the MPS soft factor before (left) and after (right) the transformation $C_{n\bar n}$ (with restrictions (\ref{Cnn:restriction})). The transverse planes at light-cone infinities transforms to the points (the correspondence is shown by the same color). The traverse plane at the light-cone origin transforms into the plane $S$ formed by intersection of light-cones that are set in the green and red points.}
\end{figure}

To apply the transformation to the soft factor geometry, we make the following restriction on the parameters
\begin{eqnarray}\label{Cnn:restriction}
a\lambda<0,\qquad\bar a\bar \lambda<0,\qquad (a\bar a)^2<\frac{1}{2 \rho^2_T},
\end{eqnarray}
where $\rho_T$ is the traverse position of the most distant (from the origin) Wilson line, i.e. $\rho^2_T=\max\{-b_i^2\}$. Then the part of the transverse plane that contains the points $b_i$, transforms into the upper branch of the surface $S$. The paths of Wilson lines transform as
\begin{eqnarray}\label{Cnn:contour_tranform}
-\bar n \sigma+b_\perp\quad &\to& \quad \bar r+\omega \,\bar v(b_\perp),
\\\nn
-n \sigma+b_\perp\quad &\to& \quad r+\omega \,v(b_\perp),
\end{eqnarray}
where $0<\omega<1$. The end-points and the directions vectors are
\begin{eqnarray}\label{Cnn:vectors}
\bar v(b)&=&\frac{1}{\lambda\bar \lambda+2 a \bar a b^2}\{-\frac{\bar \lambda}{2 a},\frac{a b^2}{\bar \lambda},b\},\qquad
\bar r=\{0^+,\frac{-1}{2\bar a\bar \lambda},0_\perp\}
\\\nn
v(b)&=&\frac{1}{\lambda\bar \lambda+2 a \bar a b^2}\{\frac{\bar a b^2}{\lambda},-\frac{\lambda}{2 \bar a},b\},\qquad
r=\{\frac{-1}{2 a \lambda},0^-,0_\perp\}.
\end{eqnarray}
The vectors $v$ and $\bar v$ are lightlike, $v^2\bar v^2=0$. The end-points of the original $\mathbf{\Phi}$ at $\sigma\to \infty (0)$ correspond to the end-points of the new Wilson line at $\omega \to 0(1)$. Therefore, the transformation $C_{n\bar n}$ transforms straight half-infinite Wilson lines $\mathbf{\Phi}_{-n}$ and $\mathbf{\Phi}_{-\bar n}$ into the straight finite Wilson lines,
\begin{eqnarray}
C_{n\bar n}\mathbf{\Phi}_{-\bar n}(b)&=&\pmb{[\bar r,S(b)]},
\\
C_{n\bar n}\mathbf{\Phi}_{-n}(b)&=&\pmb{[r,S(b)]}.
\end{eqnarray}
Correspondingly, the MPS soft factor under the action of $C_{n\bar n}$ turns into the soft factor localized in the compact domain of the space-time,
\begin{eqnarray}
C_{n\bar n}\mathbf{\Sigma}(\{b\})=\mathbf{\Omega}(\{v(b),\bar v(b)\}),
\end{eqnarray}
where
\begin{eqnarray}
\Omega^{\{a_N...a_1\},\{d_N...d_1\}}(\{v,\bar v\})=\langle 0| 
T\{
\([r,S(b_N)][S(b_N),\bar r]\)^{a_Nd_N}...
\([r,S(b_1)][S(b_1), \bar r]\)^{a_1d_1}
\}
|0\rangle.
\end{eqnarray}
The graphical representation of the transformed soft factor is given in fig.\ref{fig:Cnn}. 

The soft factor $\mathbf{\Omega}$ has only UV divergences. So, we conclude that the rapidity divergences of the original soft factor $\mathbf{\Sigma}$ turn into the UV divergences of $\mathbf{\Omega}$. Such transmutation of divergences is a known feature of conformal transformation, and it can be used to relate different aspects of the theory. Probably the most known example is the relation of the BK/JIMWLK kernel to the BMS kernel \cite{Hatta:2008st} at LO. Another example is the correspondence between rapidity and soft anomalous dimensions (which is discussed in sec.\ref{sec:corespondance} in details) shown in ref.\cite{Vladimirov:2016dll}. In both references the used transformation is ${\mathcal{C}_{\bar n}(\lambda=0,a=2^{-1/2})}$. This transformation moves the transverse plane $(n)_\perp^0$ to the light-cone infinity. It precisely corresponds the relation between BK and BMS geometries, but rather disadvantageous for the TMD (and similar) soft factors because it locates a part of $\mathbf{\Sigma}$ at the light-cone infinity.

\subsection{RTRD for Drell-Yan-like soft factors}
\label{sec:RTRD}

The renormalization theorem for rapidity divergences (RTRD) for the (color-singlet singlet entries of) DY-like MPS soft factor $\mathbf{\Sigma}(\{b\})$ reads:

\textit{
The rapidity divergences associated with different directions in the MPS soft factor can be factorized from each other. At any finite order of the perturbation theory there exist the rapidity divergence renormalization factor $\mathbf{R}_n$, which contains the rapidity singularities associated with the $(-n)_\perp^\infty$, such that the rapidity renormalized soft factor
\begin{eqnarray}\label{RTRD}
\mathbf{\Sigma}^R(\{b\},\nu^+,\nu^-)=\mathbf{R}_{n}(\{b\},\nu^+)\mathbf{\Sigma}(\{b\})\mathbf{R}^\dagger_{\bar n}(\{b\},\nu^-),
\end{eqnarray}
is free from rapidity divergences.
}

The variables $\nu^\pm$ in (\ref{RTRD}) are the scales of the rapidity renormalization. The proof of RTRD is split into two parts. The first part is to prove RTRD in a conformal field theory. The second part is to extend it to QCD.

To prove RTRD in the conformal field theory we are going to use the relation between soft factors $\mathbf{\Omega}$ and $\mathbf{\Sigma}$. These soft factors are related by the conformal transformation $C_{n\bar n}$ and hence their color-singlet parts (or in other words gauge invariant parts) equal each other in the conformal field theory.  The soft factor $\mathbf{\Omega}$ has only UV divergences at cusps and multi-cusps which can be renormalized individually. Therefore, to proof RTRD in conformal field theory, it is enough to find the correspondence between divergences of soft factors, and proof that they do not mix under the transformation $C_{n\bar n}$. Then the statement of the theorem is equivalent to the statement on the existence of the renormalization of Wilson lines \cite{Dotsenko:1979wb,Brandt:1981kf}.

To associate the divergences of $\mathbf{\Sigma}$ to the divergences $\mathbf{\Omega}$ we make a geometrical deformation of $\mathbf{\Sigma}$. The deformation parameter that regularizes a particular divergence in one soft factor regularizes its analog in another soft factor. Clearly, it could be cumbersome to trace the transformation of divergences on the level of the diagrams, since the conformal transformation also affects the gauge-fixation condition.

There are UV and rapidity divergences in $\mathbf{\Sigma}$. To start with, we consider the UV divergence of $\mathbf{\Sigma}$, that appears at the cusp located at $b_i$. To regularize it we perform a tiny displacement (in the transverse direction) of the end point of $\mathbf{\Phi}_{-n}(b_i)\to\mathbf{\Phi}_{-n}(b_i+\delta b)$, but leave $\mathbf{\Phi}_{-\bar n}(b_i)$ unchanged. The parameter $\delta b$ regularizes only the UV divergence at the cusp located at $b_i$, and does not affect any other divergences. In the soft factor $\mathbf{\Omega}$ it leads to the displacement of the end-point for Wilson line $\pmb{[r,S(b_i)]}\to \pmb{[r,S(b_i+\delta b)]}$, and thus regularizes the UV divergence of the cusp located at $S(b_i)$. Therefore, each cusp UV singularity of $\mathbf{\Sigma}$ maps to the cusp UV singularity of $\mathbf{\Omega}$.  

To regularize the rapidity divergences, the Wilson lines $\mathbf{\Phi}_{-n}$ should be deformed\footnote{The regularizations of rapidity divergences of non-geometrical type cannot be consider directly, because typically, such regularizations explicitly violate conformal symmetry.} away from the plane $(-n)_\perp^\infty$. There are three alternative ways to do so. For clarity we present all of them. 
\begin{itemize}
\item[] \textit{(i)} The half-infinite Wilson lines $\mathbf{\Phi}$ could be cut at a large distance $L$, preventing their intersection with  $(-n)_\perp^\infty$. It corresponds to the restriction $0<\sigma<L$ in the parameterization of contours. In the transformed soft factor, this deformation turns into the restriction $cL^{-1}<\omega<1$ on the contour parameterization (\ref{Cnn:contour_tranform}), where $c$ is a constant. Therefore, the Wilson lines do not reach the point $r$ but stop at the sphere with radius $\sim L^{-1}$ which surround this point.

\item[] \textit{(ii)} The directions of Wilson lines can be tilt from the light-cone infinitesimally \cite{Collins:2011zzd}. E.g. for the Wilson lines $\mathbf{\Phi}_{-\bar n}$ we change $\bar n\to\{1^+,-\alpha,0_\perp\}$, where $\alpha\to 0$. Then the vector $\bar r$ gains the infinitesimal\footnote{It is important to perform the limit $\alpha\to 0$ prior to the limit $\sigma\to \infty$. I.e. to keep the deviation from the light-cone infinitesimal even at the light-cone infinity. If this requirement is not satisfied, then both points $r$ and $\bar r$ turn to $\{0^+,0^-,0_\perp\}$. In this case the rapidity divergences are not factorizable.} addition $\alpha \,\delta \bar r(b)$. Thus the Wilson lines do not intersect at the point $\bar r$.

\item[] \textit{(iii)} The end-points of Wilson lines can be pushed away from $(-n)_\perp^\infty$ by shifting rooting positions outside of the transverse plane ${\mathbf{\Phi}_{-n}\mathbf{\Phi}^\dagger_{-\bar n}(b_i)\to\mathbf{\Phi}_{-n}\mathbf{\Phi}^\dagger_{-\bar n}(b_i+n b_i^-+\bar n b_i^+)}$ \cite{Li:2016axz}. In order to prevent the formation of another $(-n)_\perp^\infty$ (with different vector $s^\mu$), all parameters $b_i^\pm$ should be different. In this case the end-points of Wilson lines in $\mathbf{\Omega}$ do not meet at $r$ and $\bar r$ but distributed along light-cone axes with coordinates $r+nb_i^-/\lambda^2$ and $\bar r+\bar n b_i^+/\bar \lambda^2$.
\end{itemize}
In all cases the Wilson lines do not join\footnote{To ensure the gauge-invariance one should add extra transverse links which connect end points. In the soft factor $\mathbf{\Omega}$ these links would turn into the curved links. There are not extra cusp UV divergences in this case since the directions of links at meeting points are perpendicular.} together at points $r$ and $\bar r$. Therefore, we conclude that rapidity divergences of $\mathbf{\Sigma}$ turn into the UV multi-cusp divergences of $\mathbf{\Omega}$. Moreover, the rapidity divergence associated with $(-n)_\perp^\infty$ ($(-\bar n)_\perp^\infty$) turns into the separate UV divergences at $r$ ($\bar r$).

The UV divergences of soft factor $\mathbf{\Omega}$ at points $r$ and $\bar r$ are removed by renormalization factors  $\mathbf{Z}(\{v\})$ and $\mathbf{Z}^\dagger(\{\bar v\})$ independently \cite{Dotsenko:1979wb,Brandt:1981kf}. In other words,
\begin{eqnarray}\label{RTRD:1}
\mathbf{\Omega}^{\text{UV-finite at $r,\bar r$}}(\{v,\bar v\},\mu,\bar \mu)=\mathbf{Z}(\{v\},\mu)\mathbf{\Omega}(\{v,\bar v\})\mathbf{Z}^\dagger(\{\bar v\},\bar\mu),
\end{eqnarray}
where $\mu$ and $\bar \mu$ are renormalization scales. Applying $C^{-1}_{n\bar n}$ to the right-hand-side we transform each factor independently and obtain the correspondence
\begin{eqnarray}\label{RTRD:2}
C^{-1}_{n\bar n}\(\mathbf{Z}(\{v\},\mu)\)=\mathbf{R}_n(\{b\},\nu^+),\qquad 
C^{-1}_{n\bar n}\(\mathbf{Z}^\dagger(\{\bar v\},\bar \mu)\)=\mathbf{R}^\dagger_{\bar n}(\{b\},\nu^-).
\end{eqnarray}
The scale $\nu^+$($\nu^-$) is a function of $\mu$($\bar \mu$). Applying inverse transformation $C^{-1}_{n\bar n}$ to the function on the left-hand-side of (\ref{RTRD:1}) we obtain the function $\mathbf{\Sigma}^R$ which is free from rapidity divergences. Therefore, the product $\mathbf{R}_{n}\mathbf{\Sigma}\mathbf{R}^\dagger_{\bar n}$ is free from rapidity divergences. 

According to the renormalization theorem, we can define a (rapidity divergence) finite rapidity anomalous dimension (RAD)
\begin{eqnarray}\label{RTRD:D_def}
\mathbf{D}(\{b\})=\frac{1}{2}\mathbf{R}_n^{-1}(\{b\},\nu^+)\nu^+\frac{d}{d\nu^+}\mathbf{R}_{n}(\{b\},\nu^+),
\end{eqnarray}
where the factor $1/2$ is set to meet the common definition of $\mathbf{D}$. The solution of this equation is
\begin{eqnarray}\label{RTRD:Rn}
\mathbf{R}_n(\{b\},\nu^+)&=&\mathbf{A}e^{-2\mathbf{D}(\{b\}) \ln(\delta^+/\nu^+)},
\end{eqnarray}
where $\mathbf{A}$ is a $\nu$-independent matrix, which represents the scheme dependant part and is set to unity in the following. The explicit form of the rapidity renormalization factor (\ref{RTRD:Rn}) together with RTRD give the explicit form of the soft factor $\mathbf{\Sigma}$. It can be written as
\begin{eqnarray}\label{RTRD:SIGMA_FAC}
\mathbf{\Sigma}(\{b\},\delta^+,\delta^-)&=&\mathbf{R}_n^{-1}(\{b\},\nu^+)\mathbf{\Sigma}_{0}(\{b\},\nu^2)(\mathbf{R}_{\bar n}^{\dagger})^{-1}(\{b\},\nu^-)
\\\nn&=&e^{2\mathbf{D}(\{b\}) \ln(\delta^+/\nu^+)}\mathbf{\Sigma}_{0}(\{b\},\nu^2)e^{2\mathbf{D}^\dagger(\{b\}) \ln(\delta^-/\nu^-)},
\end{eqnarray}
where $\nu^2=\nu^+\nu^-$, $\delta^+$($\delta^-$) represents the regulator of rapidity divergences coupled to the  scale $\nu^+$ ($\nu^-$), and the matrix $\mathbf{\Sigma}_0$ is a rapidity divergent free matrix.  The equation (\ref{RTRD:SIGMA_FAC}) is an alternative form of RTRD (\ref{RTRD}). Although it is written in the $\delta$-regularization, it can be written in any rapidity regulator by replacing $\ln\delta$ by the corresponding rapidity divergent function.

The subscripts $n$ and $\bar n$ on the normalization factors $\mathbf{R}$ label the type of rapidity divergences (and hence the regulator), which are collected in the factors. The renormalization scales $\nu^\pm$ are not boost invariant, but transforms as corresponding components of a vector. It can be seen by considering an effect of the rescaling of geometrical regulators onto parameters $\nu$. The function $\mathbf{\Sigma}_0$ depends only on the product of $\nu^2=\nu^+\nu^-$ in the consequence of Lorentz invariance.

Next, we promote the theorem to QCD. We start with the consideration of QCD in the critical regime, where its conformal invariance is restored, and hence the equation (\ref{RTRD:SIGMA_FAC}) holds. There are several possibilities to turn QCD to the critical regime in the perturbation theory, see e.g. \cite{Braun:2014vba,Banks:1981nn}. We found it convenient to use the critical number of space-time dimension, $d^*=4-2\epsilon^*$. The value of $\epsilon^*$ is determined by the relation $\beta(\epsilon^*)=0$ order-by-order in the perturbation theory. Using the expression for the $\beta$-function in the dimensional regularization we find
\begin{eqnarray}\label{RTRD:e^*}
\epsilon^*=-a_s \beta_0-a_s^2 \beta_1-a_s^3\beta_2-...~.
\end{eqnarray}
Note, that the UV divergences of $\mathbf{\Omega}$ at $r$ and $\bar r$ should be regularized by a non-dimensional regulator (e.g. by the cut of $\omega$). At the critical number of space-time dimension, the theorem holds up to an arbitrary order of the perturbation theory. The physical QCD is defined at $\epsilon=0$. To obtain the theorem in the physical QCD we push the $\epsilon^*$ to the $0$ order-by-order in the perturbation theory. So, at the first step the $\epsilon^*$ is shifted by $\epsilon^*\to \epsilon^{**}+\beta_0 a_s$. Since QCD is conformal invariant at one-loop level, and the counting of rapidity divergences is not affected by dimensional regularization, the form of the soft factor (\ref{RTRD:SIGMA_FAC}) is preserved, with slightly changed values of $\mathbf{D}$ and $\mathbf{\Sigma}_0$. Such shift can be repeated $K$ times, with increasing perturbative order. This defines constants $\mathbf{R}$ at $(0+\mathcal{O}(a_s^{K+1}))$-number of dimension. Alternatively, this statements can be checked by solving the renormalization group equation order-by-order in a shift parameter. Thus, we have proved the theorem in conformal theory and at arbitrary order of QCD perturbation theory. 

So far we do not specify the renormalization factors for cusps. The cusp renormalization can be done before or after the rapidity renormalization. The order of renormalization affects the value of $\mathbf{R}$, due to the presence of double poles. These double poles have a geometrical origin, see e.g. \cite{Korchemskaya:1992je,Drummond:2007aua}, and do not influence the combinatorics of the subtractions. However, the order influences the relative compositions of renormalization scales. So the completely renormalized soft factor takes form
\begin{eqnarray}
\mathbf{\Sigma}^{R,R}(\{b\},\nu^+,\nu^-,\mu)=\prod_{i=1}^N Z_{i,\text{cusp}}(\mu)
\mathbf{R}^\dagger_{n}(\{b\},\nu^+)\mathbf{\Sigma}(\{b\})\mathbf{R}_{\bar n}(\{b\},\nu^-).
\end{eqnarray}
Since the renormalization factors $Z$ are scalars, it is more convenient to present RTRD in the symmetric form combining the singular factors together 
\begin{eqnarray}\label{RDRT:sym_rep}
\mathbf{\Sigma}^{R,R}(\{b\},\zeta,\bar \zeta,\mu)=\(\prod_{i=1}^N Z^{1/2}_{i,\text{cusp}}(\mu)
\mathbf{R}^\dagger_{n}(\{b\},\nu^+)\)\mathbf{\Sigma}(\{b\})\(\mathbf{R}_{\bar n}(\{b\},\nu^-)\prod_{i=1}^N Z^{1/2}_{i,\text{cusp}}(\mu)\).
\end{eqnarray}
The equation (\ref{RTRD:SIGMA_FAC}) transforms into
\begin{eqnarray}\label{RTRD:SIGMA_FAC_Z}
&&\mathbf{\Sigma}(\{b\})=
\\\nn&&\prod_{i=1}^N Z^{1/2}_{i,\text{cusp}}(\mu) e^{2\mathbf{D}(\{b\},\mu) \ln(\delta^+/\nu^+)}\mathbf{\Sigma}_{0}(\{b\},\nu^2,\mu)e^{2\mathbf{D}^\dagger(\{b\},\mu) \ln(\delta^-/\nu^-)}Z^{1/2}_{i,\text{cusp}}(\mu).
\end{eqnarray}
The $\mu$-dependence of RAD can be found by combining equations (\ref{RTRD:D_def}), (\ref{RTRD:SIGMA_FAC}), and (\ref{RDRT:sym_rep}) into
\begin{eqnarray}\label{RTRD:3}
\mu^2\frac{d}{d\mu^2} e^{2\mathbf{D}(\{b\},\mu) \ln(\delta^+/\nu^+)}=\frac{1}{4}\sum_{i=1}^N\Gamma^i_{\text{cusp}}e^{2\mathbf{D}(\{b\},\mu) \ln(\delta^+/\nu^+)},
\end{eqnarray}
where $\Gamma^i_{\text{cusp}}$ is 
\begin{eqnarray}
\Gamma^i_{\text{cusp}}=(Z^{i}_{\text{cusp}})^{-1}\mu\frac{d}{d\mu}Z^i_{\text{cusp}}.
\end{eqnarray}
The equation (\ref{RTRD:3}) can be written in the convenient form
\begin{eqnarray}\label{RTRD:RGE}
\mu^2 \frac{d}{d\mu^2}\mathbf{D}(\{b\},\mu)=\frac{1}{4}\sum_{i=1}^N\Gamma^i_{\text{cusp}}\mathbf{I}.
\end{eqnarray}
This is the generalization of the well-known Collins-Soper (CS) equation \cite{Collins:1984kg} to the MPS case. Note, that the values of $\Gamma^i_{\text{cusp}}$ differ only by color factors, since the angles of all cusps are the same. In the scalar case $N=2$, which corresponds to the TMD RAD, the color representation of both cusps are the same. In this case the equation (\ref{RTRD:RGE}) is reduced to the original CS equation \cite{Collins:1984kg}
\begin{eqnarray}\label{TMD:D_RGE}
\mu^2 \frac{d}{d\mu^2}\mathcal{D}^i(b,\mu)=\frac{\Gamma^i_{\text{cusp}}}{2}.
\end{eqnarray} 
For $N=4$ it has been checked in \cite{Vladimirov:2016qkd} at NNLO (see also discussion in sec.\ref{sec:non-dipole}).

Let us mention that there is also a possibility to leave the UV divergences unrenormalized since practically the soft factor is always combined with parton distributions. The obtained combination can be renormalized as a whole. This approach requires less algebra and thus is more convenient practically. For example, it has been used in \cite{Echevarria:2015usa,Echevarria:2016scs} for NNLO calculations.

\section{Some consequences and extensions}
\label{sec:consiquences}

\subsection[MPS factorization]{MPS factorization\footnote{I thank M.Diehl for the help in the elaboration of consistent definitions presented in this section.}}
\label{sec:MPS_fac}

The RTRD states that the rapidity divergences related to different directions are factorizable. Thus, we can finalize the factorization theorem for the multi-DY process and define a divergence-free multi-parton distribution (multiPD). Note, that all expressions presented in this section are easily reduced to the case of TMD factorization. To obtain the TMD expressions, one should only remove the $\{\}$-brackets from variables and release the color structure (see also sec.\ref{sec:scheme}).

The multi-DY scattering is characterized by momenta of produced hard particles $q_i$, with $q_i^2=Q_i^2+q_{iT}^2$. In the regime $Q_i^2\gg q_{Ti}^2$ the hadron tensor of the MPS has can be written in the factorized form \cite{Diehl:2011yj,Diehl:2015bca}
\begin{eqnarray}\label{MDY:cross}
W(\{q\})&=&\prod_{i=1}^{N/2} \sum_{f,\bar f} H_{i,f_i\bar f_i} \(\frac{Q^2_i}{\mu^2}\)
\\\nn && \int d^2 b_i d^2 b_{N-i}e^{-i (q_{i}\cdot (b_i-b_{N-i}))_\perp}
\tilde F^T_{\{\bar f\}\ot h_2}(\{\bar x\},\{b\},\mu)
\times
\mathbf{\Sigma}(\{b\},\mu)
 \times 
\tilde F_{\{f\}\ot h_1}(\{x\},\{b\},\mu),
\end{eqnarray}
where $H$ are hard scattering coefficient functions, $x$ and $\bar x$ are Bjorken variables, $\mu$ is a common hard-factorization scale $\mu\sim Q_{1,..,N/2}$. The multiPD is given by the following matrix element
\begin{eqnarray}\label{MDY:MPD}
&&\tilde F_{\{f\}\ot h}(\{x\},\{b\})=\int \Big(\prod_{i=1}^{N/2} \frac{dy_i^-dy_{N-i}^-}{(2\pi)^2} e^{i x_i (y^-_i-y^-_{N-i}) p^+}\Big)
\\&&\nn
\langle h|\bar T\{\bar \xi_{f_1}(y_1^-,b_1)
...\bar \xi_{f_{N/2}}(y_{N/2}^-,b_{N/2})\}T\{
\xi_{f_{N/2+1}}(y_{N/2+1}^-,b_{N/2+1})
...\xi_{f_{N}}(y_{N}^-,b_{N})
\}|h \rangle.
\end{eqnarray}
The Lorentz structure of multiPDs is omitted for simplicity. In both formulas, a single variable $b_i$ and a single variable $y_i^-$ can be set to zero by the translation invariance, and corresponding integrals eliminated. The fields $\xi$ can be quark, anti-quark and gluon fields with adjusted half-infinite line Wilson lines, e.g. $\xi_{q}(x)=\mathbf{\Phi}_{-n}(x)q(x)$. Therefore, the multi-parton distribution $\tilde F$ is the vector in the color space. Consequently, the multiPD $\tilde F^T$ is a row in the color space. The multiPDs $F$ are non-zero only for a color singlet combinations of indices. It automatically eliminates the non-gauge-invariant parts of the soft factor $\mathbf{\Sigma}$.

The fields participated in the definition (\ref{MDY:MPD}) are collinear fields. It implies that the soft modes of these fields should be subtracted (so-called zero-bin subtractions, see e.g\cite{Becher:2014oda}). The procedure of subtraction is dependent on the rapidity-divergences regularization. In the convenient regulator, it can be presented by an inverse soft factor see e.g.\cite{Echevarria:2016scs}  (or product of soft factors, see e.g. \cite{Collins:2011ca}). Till the end of this section we use the $\delta$-regularization for explicitness. However, the derivation can be performed in any other regularization scheme in the same manner and with the same final result. In the $\delta$-regularization, the zero-bin subtraction take the form of the inverse soft factor \cite{GarciaEchevarria:2011rb,Echevarria:2016scs}
\begin{eqnarray}\label{MDY:F-Fus}
\tilde F_{\{f\}\ot h_1}(\{x\},\{b\},\mu,\delta^-)=\mathbf{\Sigma}^{-1}(\mu;\delta^+,\delta^-)\times \tilde F^{\text{us}}_{\{f\}\ot h_1}(\{x\},\{b\},\mu,\delta^+),
\end{eqnarray}
where $\tilde F^{\text{us}}$ is the unsubtracted multiPD, i.e. evaluated directly as it stands in (\ref{MDY:MPD}).

The factorization theorem (\ref{MDY:cross}) is not complete in the sense that it does not express the cross-section via finite quantities, which depend only on a single hadron. The problem here is rapidity divergences which are presented in every constituent of the theorem. The multiPD $F$ ($F^T$) has rapidity divergences due to the interaction of far end points of Wilson lines, i.e. divergences are localized at $(-n)_\perp^\infty$ ($(-\bar n)_\perp^\infty$), and regularized by $\delta^+$ ($\delta^-$). The rapidity divergences cancel in the product $F^T(\delta^+)\times \mathbf{\Sigma}(\delta^+,\delta^-)\times F(\delta^-)$ by the statement of the factorization theorem. To complete the factorization theorem, we apply RTRD, and insert the soft factor in the form (\ref{RTRD:SIGMA_FAC}). Since the multiPD $F_{\{f\}\ot h_1}$ contains only rapidity divergences regularized by $\delta^-$ the following combination is free from rapidity divergences
\begin{eqnarray}\label{MDY:F_finite}
F_{\{f\}\ot h_1}(\{x\},\{b\},\nu^+)&=&\mathbf{\Sigma}_0(\nu^2) \mathbf{R}_{\bar n}^{\dagger\,-1}(\{b\},\nu^-)\times \tilde F_{\{f\}\ot h}(\{x\},\{b\},\delta^-)
\\\nn &=&\mathbf{\Sigma}_0(\nu^2)e^{2\mathbf{D}^\dagger(\{b\}) \ln(\delta^-/\nu^-)}\times \tilde F_{\{f\}\ot h}(\{x\},\{b\},\delta^-),
\end{eqnarray}
where the finite prefactor $\mathbf{\Sigma}_0$ is put for the future convenience. Note, that the left-hand-side of this equation is independent on $\nu^-$. It became explicit in the terms of unsubtracted multiPDs (\ref{MDY:F-Fus}), where
\begin{eqnarray}\label{MDY:F_finite_us}
F_{\{f\}\ot h_1}(\{x\},\{b\},\nu^+)&=&\mathbf{R}_n(\{b\},\nu^+)\times \tilde F^{\text{us}}_{\{f\}\ot h}(\{x\},\{b\},\delta^+)
\\\nn &=&e^{-2\mathbf{D}(\{b\}) \ln(\delta^+/\nu^+)}\times \tilde F^{\text{us}}_{\{f\}\ot h}(\{x\},\{b\},\delta^+).
\end{eqnarray}
Making the similar redefinition of $\tilde F^T$ we obtain the factorization theorem in the form
\begin{eqnarray}\nn
&&\tilde F^T(\{\bar x\},\{b\},\mu)\times \mathbf{\Sigma}(\{b\},\mu)\times \tilde F^T(\{x\},\{b\},\mu)=
\\&&\label{MDY:rap_FAC}
\qquad\qquad\qquad\qquad\qquad
F^T(\{\bar x\},\{b\},\mu,\nu^-)\times \mathbf{\Sigma}^{-1}_0(\{b\},\mu,\nu^2)\times F(\{x\},\{b\},\mu,\nu^+).
\end{eqnarray}
Here all components are finite. And thus, the factorization theorem is completed.

The dependence of a multiPD on the rapidity scales follows from the equations (\ref{MDY:F_finite_us}) and (\ref{RTRD:D_def}),
\begin{eqnarray}\label{MDY:nu-evol}
\nu^+ \frac{d}{d\nu^+}F_{\{f\}\ot h}(\{x\},\{b\},\mu,\nu^+)=\frac{1}{2}\mathbf{D}^{\{f\}}(\{b\},\mu)\times F_{\{f\}\ot h}(\{x\},\{b\},\mu,\nu^+).
\end{eqnarray}
The factorized expression (\ref{MDY:rap_FAC}) contains the multiPDs that depend on the variables $\nu^+$ and $\nu^-$, which seems to contradict the Lorentz invariance. Nonetheless, there is no contradiction, because the multiPDs are defined on the states with momenta oriented along $n$ or $\bar n$. It allows to pass to a more convenient (and standard) boost invariant variables $\zeta$ and $\bar \zeta$, which is done in the next section.

We also note that the rapidity divergences are independent on the kind of states. They are the part of the operator, similarly to UV divergences. Therefore, the factor $\mathbf{R}$ applies directly to the multiPD operator. Such composition greatly  simplifies the study of properties of multiPD operators without reference to the parton model consistently. For example, to perform the operator product expansion in the background field technique.

\subsubsection{Boost invariant variables and scheme dependence}
\label{sec:scheme}

Let us introduce the boost invariant variables
\begin{eqnarray}\label{zeta-def}
\zeta=2(p^+)^2\frac{\nu^-}{\nu^+},\qquad \bar \zeta=2(p^-)^2\frac{\nu^+}{\nu^-},\qquad \zeta\bar \zeta=(2p^+p^-)^2
\end{eqnarray}
where $p^+$ and $p^-$ are components of a vector $p^\mu$. Vector $p^\mu$ can be selected arbitrary, but it is convenient to associate it with the momentum of the produced particle (e.g. with the momentum of the produced photon for the DY processes). In this case, we have $\zeta\bar \zeta=Q^4$ where $Q$ is the typical virtuality of the process. Assuming this, the multiPD becomes a function of $\zeta$ and $\nu^2$, i.e. $F(\{x\},\{b\},\mu,\zeta,\nu^2)$. The $\zeta$ dependence follows from equation (\ref{MDY:nu-evol}),
\begin{eqnarray}
\zeta \frac{d}{d\zeta}F_{\{f\}\ot h}(\{x\},\{b\},\mu,\zeta,\nu^2)=-\mathbf{D}^{\{f\}}(\{b\},\mu)\times F_{\{f\}\ot h}(\{x\},\{b\},\mu,\zeta,\nu^2).
\end{eqnarray}
This equation coincides with the standard definition of the rapidity evolution (see e.g.\cite{Vladimirov:2016qkd,Diehl:2015bca,Aybat:2011zv}).

In the presented above construction defers from usual constructions, e.g. in refs.\cite{Echevarria:2016scs,Vladimirov:2016qkd,Diehl:2011yj,Diehl:2015bca}, by the presence of an extra parameter $\nu^2$. This parameter decouples from the equations and, therefore, is unrestricted. We stress that it also appears in the remnant of the soft factor $\mathbf{\Sigma}_0(\nu^2)$, which is scheme dependent. In this way, the parameter $\nu^2$ is a part of the scheme definition. 

We recall that the rapidity renormalization factors $\mathbf{R}$ are defined up to an arbitrary matrix, see (\ref{RTRD:Rn}). Therefore, the definition of the multiPD is not unique. We can introduce an alternative multiPD with the multiplication by an arbitrary finite matrix $\mathbf{S}$, i.e.
\begin{eqnarray}\label{MDY:eqeq1}
F_{\{f\}\ot h_1}(\{x\},\{b\},\zeta,\nu^2)\to \mathbf{S}\times F_{\{f\}\ot h_1}(\{x\},\{b\},\zeta,\nu^2).
\end{eqnarray}
 Such procedure does not damage the factorization theorem (\ref{MDY:rap_FAC}), and leads only to the replacement
\begin{eqnarray}\label{MDY:eqeq2}
\mathbf{\Sigma}^{-1}_0(\{b\},\mu,\nu^2)\to (\mathbf{S}^{-1})^{T}\mathbf{\Sigma}^{-1}_0(\{b\},\mu,\nu^2) \mathbf{S}^{-1}.
\end{eqnarray}
Compare equations (\ref{MDY:eqeq1},\ref{MDY:eqeq2},\ref{MDY:F_finite_us}) and (\ref{RTRD:Rn}) we conclude that the the matrix $\mathbf{S}$ can be recasted to the matrix $\mathbf{A}$, and thus, is a part of scheme definition. Since the matrix $\mathbf{S}$ is a part of the rapidity renormalization factor is can depend on any variables except $\zeta$.

The expression for matrix $\mathbf{S}$ should be fixed conveniently by some regularization-independent condition. Let us discuss the fixation of the scheme in the TMD case. The conventional form of the TMD factorization theorem (see e.g. \cite{Collins:2011zzd,Collins:2011ca,GarciaEchevarria:2011rb,Echevarria:2012js,Chiu:2012ir,Echevarria:2015byo,Echevarria:2016scs,Luebbert:2016itl,Li:2016axz}) defines the hadron tensor as a product of two TMD distributions without any remnant of the soft factor matrix $\Sigma_0$. The TMD hadron tensor reads
\begin{eqnarray}\label{TMD:fffff}
W_{\text{TMD}}&=&\sum_{\bar f,f}H_{\bar f f}\(\frac{Q^2}{\mu^2}\)\int \frac{d^2b}{(2\pi)^2} e^{i (q\cdot b)_T}
F_{\bar f\ot h_2}(\bar x,b,\mu,\bar \zeta)
F_{f\ot h_1}(x,b,\mu,\zeta).
\end{eqnarray}
This form of the factorization theorem agrees with the parton model picture, since the hard coefficient can be interpreted as the cross-section of parton scattering, and at small-$b$ $F(x,b\to 0)\to f(x)$, where $f(x)$ is the usual parton distribution function. The expression (\ref{TMD:fffff}) implies the following relation
\begin{eqnarray}\label{TMD:scheme}
S^{-1}(b,\mu, \nu^2)\Sigma_0^{-1\,\text{TMD}}(b,\mu, \nu^2)S^{-1}(b,\mu,\nu^2)=1.
\end{eqnarray}
Using this scheme we obtain the following expression for TMD distribution
\begin{eqnarray}
F_{f\ot h}(x,b,\mu,\zeta,\nu^2)&=&\sqrt{\Sigma_0^{\text{TMD}}(b,\mu,\nu^2)}e^{-2\mathcal{D}^f(b,\mu)\ln(\delta^+/\nu^+)}\tilde F^{\text{us}}_{f\ot h}(x,b,\delta^+).
\end{eqnarray}
Recalling the simple structure of the TMD soft factor (\ref{TMD:structure}) we arrive to the standard expression for the TMD distribution
\begin{eqnarray}
F_{f\ot h}(x,b,\mu,\zeta)&=&\sqrt{\Sigma_{\text{TMD}}\(b,\frac{\delta^+}{\sqrt{2}p^+}\sqrt{\zeta},\frac{\delta^+}{\sqrt{2}p^+}\sqrt{\zeta}\)}\tilde F_{f\ot h}(x,b,\delta^+).
\end{eqnarray}
The $\nu^2$ parameter is not presented in this definition.

The natural generalization of the TMD scheme fixation condition (\ref{TMD:scheme}) for the MPS case is
\begin{eqnarray}
\mathbf{S}(b,\mu,\nu^2)\mathbf{\Sigma}_0(\{b\},\mu,\nu^2) \mathbf{S}^T(b,\mu,\nu^2)=\mathbf{I}.
\end{eqnarray}
In the recent paper \cite{Buffing:2017mqm}, it has been shown that in the $N=4$ case the solution of this equation exists and naturally expresses in the terms of matrices $s$ (\ref{DPD:small_s}). In this scheme the MPS factorization theorem is 
\begin{eqnarray}\label{MDY:cross2}
W(\{q\})&=&\prod_{i=1}^{N/2} \sum_{f,\bar f} H_{i,f_i\bar f_i} \(\frac{Q^2_i}{\mu^2}\)
\\\nn && \int d^2 b_i d^2 b_{N-i}e^{-i (q_{i}\cdot (b_i-b_{N-i}))_\perp}
F^T_{\{\bar f\}\ot h_2}(\{\bar x\},\{b\},\mu,\zeta)
\times
F_{\{f\}\ot h_1}(\{x\},\{b\},\mu,\zeta).
\end{eqnarray}
Such scheme is equivalent to the decomposition of the soft factor (\ref{DPD:FAC}). In the case of DPDs this decomposition has been explicitly demonstrated at NNLO in \cite{Vladimirov:2016qkd}. Note, that generally speaking the matrix $\mathbf{S}$ does not commute with $\mathbf{D}$ and therefore, the rapidity anomalous dimension is scheme dependent
\begin{eqnarray}
\mathbf{D}_{S}=\mathbf{S}\mathbf{D}\mathbf{S}^{-1}\sim \mathbf{D}+a_s[\mathbf{s},\mathbf{D}]+\mathcal{O}(a_s^2),
\end{eqnarray}
where for the last equality we substitute $\mathbf{S}=\exp(a_s\mathbf{s})$. The explicit evaluation of color structure presented in sec.\ref{sec:color} shows that $[\mathbf{s},\mathbf{D}]\sim \mathcal{O}(a_s^3)$ at least.

\subsection{Correspondence between soft and rapidity anomalous dimensions}
\label{sec:corespondance}

The relation between the rapidity and UV singularities give rise to the correspondence between RAD and SAD \cite{Vladimirov:2016dll}. The correspondences between anomalous dimensions are highly interesting, since they connect different regimes of physics. To our best knowledge, nowadays there are only two examples of such correspondences in QCD: the discussed here SAD-to-RAD correspondence, and the BK/JIMWLK-to-BMS correspondence \cite{Hatta:2008st}. The check of SAD-to-RAD correspondence gives a non-trivial confirmation of RTRD.

The soft anomalous dimension (SAD) is defined as
\begin{eqnarray}
\pmb{\gamma}_s(\{v\})=\mathbf{Z}^{-1}(\{v\},\mu)\mu\frac{d}{d\mu}\mathbf{Z}(\{v\},\mu),
\end{eqnarray}
where $\mathbf{Z}$ is the UV renormalization factor for multi-cups non-analyticity of Wilson lines, that appear in $\mathbf{\Omega}$ (\ref{RTRD:1}). Comparing to (\ref{RTRD:D_def}) we obtain the exact relation in the conformal field theory
\begin{eqnarray}\label{RAD-SAD:conformal}
\pmb{\gamma}_s(\{v\})=2\mathbf{D}(\{b\}),
\end{eqnarray}
where vectors $v$ and $b$ are related by $C_{n\bar n}$ transformation. This relation has been observed for the TMD case in the conformal invariant $\mathcal{N}=4$ super-Yang-Mills theory at three-loop order \cite{Li:2016ctv}.

In QCD the equality (\ref{RAD-SAD:conformal}) holds at the critical point (\ref{RTRD:e^*}). The UV anomalous dimension is $\epsilon$-independent, in the contrast to the RAD. Therefore, we have
\begin{eqnarray}\label{RAD-SAD}
\pmb{\gamma}_s(\{v\})=2\mathbf{D}(\{b\},\epsilon^*).
\end{eqnarray}
Using this expression the physical value of RAD (SAD) can be obtained at a given perturbative order using the finite part of the previous perturbative order and the know expression for SAD (RAD). Indeed, substituting $\epsilon^*$ in the form (\ref{RTRD:e^*}) and comparing the coefficients for different powers of $a_s$ we obtain
\begin{eqnarray}\label{RAD-SAD:1}
\mathbf{D}_1(\{b\})&=&\frac{1}{2}\pmb{\gamma}_1(\{v\}),
\\\label{RAD-SAD:2}
\mathbf{D}_2(\{b\})&=&\frac{1}{2}\pmb{\gamma}_2(\{v\})+\beta_0 \mathbf{D}'_1(\{b\}),
\\\label{RAD-SAD:3}
\mathbf{D}_3(\{b\})&=&\frac{1}{2}\pmb{\gamma}_3(\{v\})+\beta_0 \mathbf{D}'_2(\{b\})+\beta_1 \mathbf{D}'_1(\{b\})-\frac{\beta_0^2}{2}\mathbf{D}''_1(\{b\}),
\end{eqnarray} 
and so on. Here, we use the notation $\pmb{\gamma}=\sum a_s^n \pmb{\gamma}_{n}$ and $\mathbf{D}=\sum a_s^n \mathbf{D}_n$, and primes denote the derivatives with respect to $\epsilon$ at $\epsilon=0$.

\subsubsection{TMD rapidity anomalous dimension at three-loop order}
\label{sec:SADRAD}

The practically most interesting case is the TMD RAD. It is corresponded to the dipole part of the SAD, or to the lightlike cusp anomalous dimension. The expression for the cusp anomalous dimension is known up to three-loop order \cite{Moch:2004pa}, which allows us to learn the three-loop RAD, using the two-loop calculation.

The dipole contribution to the SAD has the form
\begin{eqnarray}
C_i\gamma_{\text{dipole}}(v_i\cdot v_j)
=\ln\(\frac{(v_i\cdot v_j)\mu^2}{\nu_{ij}^2}\)\Gamma^i_{cusp}-\tilde \gamma^i_s,
\end{eqnarray}
where $\nu_{ij}^2$ is a IR scale which regularizes the lightlike cusp angle, and $C_i$ is the quadratic Casimir eigenvalue. The coefficients of the perturbative expansion for $\Gamma$ and $\gamma_s$ can be found in \cite{Moch:2004pa}, and are given in the appendix \ref{app:ADs}. The NLO TMD anomalous dimension at arbitrary $\epsilon$ is \cite{Echevarria:2015byo} (see also (\ref{SF:1loop}))
\begin{eqnarray}\label{RAD-SAD:D1}
\mathcal{D}_1^i(b,\epsilon)&=&-2a_sC_i \(B^\epsilon\Gamma(-\epsilon)+\frac{1}{\epsilon}\),
\end{eqnarray}
where $B=b^2\mu^2/4 e^{-2\gamma_E}$. Using the equation (\ref{RAD-SAD:1}) we obtain the equality
\begin{eqnarray}
\frac{\gamma_{1,\text{dipole}}}{2}=2\ln\(\frac{(v_1\cdot v_2)\mu^2}{\nu_{12}^2}\)=\frac{\mathcal{D}^i_1}{C_i}=2\ln\(\frac{b_{12} \mu^2}{4 e^{-2\gamma_E}}\).
\end{eqnarray}
The vectors $v$ and $b$ are related by (\ref{Cnn:vectors}) which gives
\begin{eqnarray}\label{RAD-SAD:transform}
C_{nn}(v_i\cdot v_j)=\frac{b_{12}}{(\lambda \bar \lambda+ a \bar a b_1^2)(\lambda \bar \lambda+ a \bar a b_2^2)}.
\end{eqnarray}
It fixes the relative scheme dependence between rapidity renormalization and UV renormalization
\begin{eqnarray}
\nu_{ij}=4 e^{-2\gamma_E}(\lambda \bar \lambda+ a \bar a b_i^2)(\lambda \bar \lambda+ a \bar a b_j^2).
\end{eqnarray}

At the order $a_s^2$,  RAD has an extra logarithm structure which is produced by the expansion of $B^\epsilon$ in (\ref{RAD-SAD:D1}). Therefore, comparing left and right hand sides of (\ref{RAD-SAD:2}) we find 
\begin{eqnarray}
\mathcal{D}_2=d^{(2,2)}L^2_b+d^{(2,1)}L_b+d^{(2,0)}=\beta_0 L_b^2+2\Gamma_1 L_b-\frac{\tilde \gamma_1}{2}+\beta_0 \zeta_2,
\end{eqnarray}
where
\begin{eqnarray}
d^{(2,2)}=\beta_0,\qquad d^{(2,1)}=2\Gamma_1,\qquad d^{(2,0)}=-\frac{\tilde \gamma_1}{2}+\beta_0\zeta_2.
\end{eqnarray}
These numbers coincide with RAD coefficients calculated directly, see e.g. \cite{Echevarria:2015byo,Becher:2010tm,Luebbert:2016itl}. 

To obtain the RAD at NNLO the $\epsilon$-dependent NLO expression is required. It has been evaluated in \cite{Echevarria:2015byo}, and reads
\begin{eqnarray}\label{RAD-SAD:D2}
\mathcal{D}_2^i(b,\epsilon)&=&2 C_i\Bigg\{ B^{2\epsilon}\Gamma^2(-\epsilon)\Big[C_A(2\psi(-2\epsilon)-2\psi(-\epsilon)+\psi(\epsilon)+\gamma_E)
\\&&\nn+\frac{1-\epsilon}{(1-2\epsilon)(3-2\epsilon)}\(\frac{3(4-3\epsilon)}{2\epsilon}C_A-N_f\)\Big] +B^\epsilon \frac{\Gamma(-\epsilon)}{\epsilon}\beta_0+\frac{\beta_0}{2\epsilon^2}-\frac{\Gamma_1}{2\epsilon}\Bigg\}.
\end{eqnarray}
Substituting it into equation (\ref{RAD-SAD:3}) we obtain
\begin{eqnarray}
\mathcal{D}_3&=&d^{(3,3)}L_b^3+d^{(3,2)}L_b^3+d^{(3,1)}L_b+d^{(3,0)}=2 \Gamma_2 L_b-\frac{\tilde \gamma_2}{2}
\\\nn&&\quad-\frac{\beta_0^2}{3}L_b^3+\beta_1 L_b^2-\beta_0^2 \zeta_2L_b-\frac{2}{3}\beta_0^2 \zeta_3+\beta_1\zeta_2
\\\nn&&\quad+\beta_0^2 L_b^3+2\beta_0\Gamma_1L_b^2+\beta_0(2d^{(2,0)}+\beta_0\zeta_2)L_b+\beta_0\Gamma_1\zeta_2+
\beta_0\[C_A\(\frac{2428}{81}-26\zeta_4\)-N_f\frac{328}{81}\],
\end{eqnarray}
where the second line comes from the expansion of $\mathcal{D}_1$ (\ref{RAD-SAD:D1}), and the third line comes from the expansion of  $\mathcal{D}_2$ (\ref{RAD-SAD:D2}). The coefficients $d^{(n,k)}$ are
\begin{eqnarray}
d^{(3,3)}&=&\frac{2}{3}\beta_0^2,\qquad d^{(3,2)}=2\Gamma_1\beta_0+\beta_1,\qquad d^{(3,1)}=2\beta_0d^{(2,0)}+2\Gamma_2,
\\\nn 
d^{(3,0)}&=&-\frac{\tilde \gamma_2}{2}+(\beta_1+\beta_0\Gamma_1) \zeta_2-\frac{2}{3}\beta_0^2 \zeta_3+\beta_0\[C_A\(\frac{2428}{81}-26\zeta_4\)-N_f\frac{328}{81}\].
\end{eqnarray}
Substituting the explicit expressions anomalous dimensions we obtain
\begin{eqnarray}
d^{(3,0)}&=&C_A^2\(\frac{297029}{1458}-\frac{3196}{81}\zeta_2-\frac{6164}{27}\zeta_3-\frac{77}{3}\zeta_4+\frac{88}{3}\zeta_2\zeta_3+96\zeta_5\)
\\\nn&&+C_AN_f\(-\frac{31313}{729}+\frac{412}{81}\zeta_2+\frac{452}{27}\zeta_3-\frac{10}{3}\zeta_4\)
\\\nn&&+C_FN_f\(-\frac{1711}{54}+\frac{152}{9}\zeta_3+8\zeta_4\)+N_f^2\(\frac{928}{729}+\frac{16}{9}\zeta_3\).
\end{eqnarray}
This expression coincides with the expression obtained in \cite{Vladimirov:2016dll,Li:2016ctv}.

The obtained expressions satisfy the renormalization group equation for TMD RAD (\ref{TMD:D_RGE}). On one hand side, it gives an extra check for the calculation. On another hand side, it is not accidental. The UV anomalous dimensions are $\epsilon$-independent by definition, and therefore the equation (\ref{RTRD:RGE}) holds at arbitrary $\epsilon$.

\subsubsection{Leading non-dipole contribution to rapidity anomalous dimension}
\label{sec:non-dipole}

The leading contributions to the non-dipole SAD has been calculated in \cite{Almelid:2015jia}. In accordance to (\ref{RAD-SAD:1}), the leading non-dipole contribution to RAD can be obtained by the direct transformation.

The leading non-dipole contribution to SAD appears at the three-loop level. The SAD at this order has the form \cite{Almelid:2015jia}
\begin{eqnarray}\label{SAD:color}
\pmb{\gamma}_s(\{v\})&=&-\frac{1}{2}\sum_{[i,j]}
\mathbf{T}^A_i\mathbf{T}^A_j \gamma_{\text{dipole}}(v_i\cdot v_j)-
\sum_{[i,j,k,l]}if^{ACE}if^{EBD}\mathbf{T}_i^A\mathbf{T}_j^B\mathbf{T}_k^C\mathbf{T}_l^D\mathcal{F}_{ijkl}
\\&&\nn
-
\sum_{[i,j,k]}\mathbf{T}_i^{\{AB\}}\mathbf{T}_j^C\mathbf{T}_k^D if^{ACE}if^{EBD}C+\mathcal{O}(a_s^4),
\end{eqnarray}
where we use the same notation as in sec.\ref{sec:color}. 

It is important that the SAD depends only on the conformal rations $\rho$ of vectors $v$ \cite{Aybat:2006mz,Gardi:2009qi}. In contrast to the transformation of the scalar product (\ref{RAD-SAD:transform}), the conformal ratios $\rho$ do not obtain any scheme factors under the transformation $C_{n\bar n}$. E.g. at N$^3$LO only the following ratios arise 
\begin{eqnarray}\label{SAD:rho->rho}
\rho_{ijkl}=\frac{(v_i\cdot v_j)(v_k\cdot v_l)}{(v_i\cdot v_k)(v_j\cdot v_l)},\qquad C_{n\bar n}(\rho_{ijkl})=\tilde \rho_{ijkl}=\frac{(b_i-b_j)^2(b_k-b_l)^2}{(b_i-b_k)^2(b_j-b_l)^2}.
\end{eqnarray}

The color structure of the MPS soft factor is elaborated in the appendix \ref{app:color} and presented in sec.\ref{sec:color}. Taking into account that the dipole part is the TMD soft factor with the structure (\ref{TMD:structure}) and the definition (\ref{RTRD:SIGMA_FAC}) we find that up to three-loop order the RAD has the following expression
\begin{eqnarray}\label{RAD:allcolor}
\mathbf{D}(\{b\})&=&-\frac{1}{2}\sum_{[i,j]}
\mathbf{T}^A_i\mathbf{T}^A_j \mathcal{D}(b_{ij})-
\sum_{[i,j,k,l]}if^{ACE}if^{EBD}\mathbf{T}_i^A\mathbf{T}_j^B\mathbf{T}_k^C\mathbf{T}_l^D\tilde{\mathcal{F}}_{ijkl}(\{b\})
\\&&\nn
-
\sum_{[i,j,k]}\mathbf{T}_i^{\{AB\}}\mathbf{T}_j^C\mathbf{T}_k^D if^{ACE}if^{EBD}\tilde C+\mathcal{O}(a_s^4).
\end{eqnarray}
The color structure literally coincides with (\ref{SAD:color}). Therefore, the functions $C$ and $\mathcal{F}$ could be obtained by replacing $\rho\to \tilde \rho$ (\ref{SAD:rho->rho}). Comparing with the parametrization of \cite{Almelid:2015jia} we obtain
\begin{eqnarray}\label{RAD:non-dipole1}
\tilde C&=&a_s^3\(\zeta_2\zeta_3+\frac{\zeta_5}{2}\)+\mathcal{O}(a_s^4),
\\\label{RAD:non-dipole2}
\tilde{\mathcal{F}}_{ijkl}(\{b\})&=&8a_s^3\mathcal{F}(\tilde \rho_{ikjl},\tilde \rho_{iljk})+\mathcal{O}(a_s^4),
\end{eqnarray}
where function $\mathcal{F}$ is given in \cite{Almelid:2015jia} in the terms of single-valued harmonic polylogarithms. 

Using the color decomposition (\ref{RAD:allcolor}) we can test the renormalization group equation (\ref{RTRD:RGE}). Differentiating (\ref{RAD:allcolor}) with respect to $\mu$ and using the renormalization group equation for the dipole part (\ref{TMD:D_RGE}) we find
\begin{eqnarray}
\mu^2 \frac{d}{d\mu^2}\mathbf{D}(\{b\})&=&-\frac{1}{2}\sum_{[i,j]}\mathbf{T}^A_i\mathbf{T}^A_j \frac{\Gamma^i_{\text{cusp}}}{2 C_i}+\mu^2\frac{d}{d\mu^2}(\text{\textbf{non-dipole terms}})
\\ \nn &=&\frac{1}{4}\sum_{i=1}^N\Gamma^i_{\text{cusp}}\mathbf{I}_i+\mu^2\frac{d}{d\mu^2}(\text{\textbf{non-dipole terms}}),
\end{eqnarray}
where the non-dipole terms include all possible non-dipole color structures starting from the leading terms presented in (\ref{RAD:allcolor}). To obtain the last line we have used the color neutrality condition (\ref{def:colorless}). Thus we conclude that at all orders of the perturbation theory
\begin{eqnarray}\label{RADSAD:nondipoleRGE}
\mu^2\frac{d}{d\mu^2}(\text{\textbf{non-dipole terms}})=0,
\end{eqnarray}
which agrees with results (\ref{RAD:non-dipole1}), (\ref{RAD:non-dipole2}).

\subsubsection{All-order constraint on the color-structure of soft anomalous dimension}

The absence of the color-tripole in the SAD is well-known. It has been shown in \cite{Aybat:2006mz,Gardi:2009qi,Dixon:2009ur}, that tripole contribution is absent at all-orders in the consequence of permutation and rescaling symmetries. Using the correspondence between SAD and RAD we can make a more restrictive statement. 

The MPS soft factor has a peculiar color-structure which follows from the generating function decomposition, see sec.\ref{sec:color} and the derivation in appendix \ref{app:color}. Namely, it has only even-color contributions at all orders (\ref{MPS:all_order_color}). The decomposition of the soft factor (\ref{RTRD:SIGMA_FAC}) does not violate such structure. It is the consequence of commutativity of generators with different indices. Indeed, commuting odd-number of generators we necessary obtain an anti-symmetric structure in some sub-set of indices, which is eliminated by the symmetric sum over all Wilson lines (the examples of such structures up to fourth order are demonstrated in appendix \ref{app:color}). Therefore, the rapidity anomalous dimension has the same color-pattern as $\mathbf{\Sigma}$,
\begin{eqnarray}
\mathbf{D}(\{b\})&=&
\sum_{\substack{n=2\\n\in\text{even}}}^{\infty}\sum_{i_1,...,i_n=1}^N\{\mathbf{T}^{A_1}_{i_1}...\mathbf{T}^{A_n}_{i_n}\}D^{n;i_1...i_n}_{A_1...A_n}(\{v\}).
\end{eqnarray}
The explicit example for $n=2$ and $n=4$ is given in (\ref{RAD:allcolor}).

The correspondence between SAD and RAD preserves the color structure. Thus, the SAD also contains only the even-number of color-generators
\begin{eqnarray}
\pmb{\gamma}_s(\{v\})&=&
\sum_{\substack{n=2\\n\in\text{even}}}^{\infty}\sum_{i_1,...,i_n=1}^N\{\mathbf{T}^{A_1}_{i_1}...\mathbf{T}^{A_n}_{i_n}\}\gamma^{n;i_1...i_n}_{A_1...A_n}(\{v\}).
\end{eqnarray}
The explicit structure involving four generators is given, e.g. in (\ref{app:MPS_colored}). The next-order color structures requires six generators, and thus appear only at fifth loop order.

\subsection{Universality of TMD soft factor}
\label{sec:TMD_UNIVERSAL}

The RTRD is formulated for the DY-like geometry of the soft factor. Such kinematics is essential, since in this case the soft factor can be written as a matrix element of a single T-ordered operator, and thus the conformal transformation could be applied. The same is true for the soft factor in the kinematics of $e^+e^-$-annihilation. In contrast, the soft factor for the SIDIS-like processes could not be analyzed in this way. However, the TMD-soft factor has a peculiarly simple structure, which leads to the equality of DY and SIDIS soft factors. Let us present it in details.

The TMD soft factor for the SIDIS kinematics reads
\begin{eqnarray}\label{SF:TMD_SIDIS}
\Sigma_{\text{TMD}}^{\text{SIDIS}}(b) = \frac{1}{N_c}
\langle 0|\bar T\{\Phi^{dc_2}_{n}(b)\Phi^{\dagger c_2a}_{-\bar n}(b)\}T\{\Phi^{ac_1}_{-\bar n}(0)\Phi^{\dagger c_1d}_{n}(0)\}  |0\rangle.
\end{eqnarray}
The fields of $\Phi_{-\bar n}$ are separated by the timelike distances from the fields of $\Phi_{n}$. Thus, one cannot present the SIDIS soft factor as a matrix element of a single T-ordered operator. 

However, the SIDIS soft factor can be factorized, as a consequence of the factorization theorem for the DY soft factor. Let us compare these soft factors within the $\delta$-regularization. On the level of Feynman diagrams the only difference between DY and SIDIS soft factors is the sign of $\delta^-$ contribution. I.e. a diagram with $n$-gluons coupled to Wilson lines $\Phi_{-\bar n}$ in the DY case has the form (in the momentum representation)
\begin{eqnarray}\label{FF:1}
I^{\text{DY}}=\int d^dk_1...d^dk_n F(\{k\},\delta^+)\frac{1}{(k^-_1+i\alpha_1\delta^-)...(k^-_n+i\alpha_n\delta^-)},
\end{eqnarray}
where $\alpha_i$ are some integers. The same diagram in the SIDIS kinematics reads
\begin{eqnarray}\label{FF:2}
I^{\text{SIDIS}}=\int d^dk_1...d^dk_n F(\{k\},\delta^+)\frac{1}{(k^-_1-i\alpha_1\delta^-)...(k^-_n-i\alpha_n\delta^-)}.
\end{eqnarray}
The function $F$ is the same in both cases. We split the integration measure as $d^dk=dk_+dk_-d^{d-2}k_\perp$, and integrate over $k^+$ components. The integration over the $k^+$ components can be done closing the integration contours on the poles of (anti-)Feynman propagators or by $\delta$-functions of cut propagators. Both cases restrict the integration over minus-components to finite or semi-infinite region of integration, $R$. Note, that contributions of eikonal poles do not restrict minus-components. Such contributions vanish in the sum of diagrams, because they result into the power-divergences in $\delta$, which necessarily cancel, see sec.\ref{sec:delta-structure}.  Therefore, the integral (\ref{FF:1}) and (\ref{FF:2}) became
\begin{eqnarray}\label{FF:3}
I^{\text{DY}(\text{SIDIS})}=\int d^{d-2}k_{1\perp}...d^{d-2}k_{n\perp}
\int dk^-_1...dk^-_n F'(\{k\},\delta^+)\frac{\theta(k_1^-,...,k_n^-\in R)}{(k^-_1\pm i\alpha_1\delta^-)...(k^-_n\pm i\alpha_n\delta^-)}.
\end{eqnarray}
In this integral, the change $\delta^-\to-\delta^-$ can be done without the crossing of the integration contour. Therefore, the SIDIS integrals are  related to the DY integral by the analytical continuation $\delta^-\to-\delta^-$. The rapidity divergences arises as $\ln(\delta^+\delta^-)$. The analytical continuation $\delta^-\to-\delta^-$ does not change the coefficient of the highest power of $\ln(\delta^+\delta^-)$, while the coefficients of lower powers can obtain extra terms proportional to $(i\pi)^k$. 

Let us note that due to the absence of color-matrix structure in the TMD case the equation (\ref{RTRD:SIGMA_FAC}) reduces to
\begin{eqnarray}
\Sigma^{\text{DY}}_{\text{TMD}}(b)=\exp\(2\mathcal{D}(b,\mu)\ln\(\frac{\delta^+\delta^-}{\mu^2}\)+B(b,\mu)\),
\end{eqnarray}
where $B$ is some rapidity divergences-free function. The logarithm contribution is not affected by analytical continuation. So the SIDIS soft factor is rapidity factorizable. The statement can be enforced. The only possible addition to the finite part $B$ should be proportional to $i\pi$. However, $\Sigma=\Sigma^\dagger$ and thus
\begin{eqnarray}
\Sigma_{\text{TMD}}^{\text{DY}}=\Sigma_{\text{TMD}}^{\text{SIDIS}}=\exp\(2\mathcal{D}(b,\mu)\ln\(\frac{|\delta^+\delta^-|}{\mu^2}\)+B(b,\mu)\).
\end{eqnarray}
This relation has been checked explicitly at NNLO in \cite{Echevarria:2015byo}. The method used here cannot be generalized to a $N>2$ case because the matrix MPS soft factor contains the higher powers of $\ln(\delta^+\delta^-)$.

\section{Conclusion}

In this work, we have considered the structure of rapidity divergences of the multi-parton scattering (MPS) soft factor. We have proven the renormalization theorem for rapidity divergences (RTRD) for MPS soft factors and discussed some of its consequences. The RTRD states that the rapidity divergences of the MPS soft factor related to different lightlike directions do not mix and can be independently renormalized. It leads to a number of consequences. The main one is the generalization of the TMD factorization theorem for a larger class of processes, e.g. double-parton scattering.

The proof of RTRD relies on the observation that the MPS soft factor can be converted to another soft factor by a conformal transformation. The obtained soft factor has a compact spatial structure and completely defined set of UV divergences. Tracing the transformation of rapidity divergences we connect the UV renormalization factor with the rapidity divergences renormalization factor. This consideration, which is valid in the conformal field theory, can be promoted to QCD using the conformal invariance of QCD at one-loop, and that the rapidity divergences are insensitive to the dimensional regularization. In this way, the RTRD can be seen as a consequence of the renormalization theorem for ultraviolet (UV) divergences and the counting rules for rapidity divergences.

We have studied the rapidity renormalization for the soft factors typical for the Drell-Yan processes. The same procedure can be done for more general soft factors. The only requirement is the possibility to rewrite the soft factor as a matrix element of single T-ordered operator. In the article, we demonstrate an example where the absence of this requirement does not destroy RTRD. This is the TMD soft factor for SIDIS. In this case, the analytical continuation between the DY and SIDIS soft factors can be performed. As a result, these soft factors are equal to each other, what has been discussed in the literature for a long time, see e.g. \cite{Collins:2011zzd,Echevarria:2015byo,Echevarria:2016scs}. A similar study is not obviously possible for many other kinematic configurations. E.g. the soft factors for processes with jets that have restrictions on the integration phase-space \cite{Stewart:2010tn,Jouttenus:2011wh}. Nonetheless, even in these cases the application of the conformal transformation $C_{n\bar n}$ (or its analogue) can give a hint on the structure of divergences.

In general, the graph-topological structure of rapidity divergences is the same as for UV divergences  (sec.\ref{sec:rap_div}). In this light, RTRD can be seen as the rule for the subdiagram subtractions, which splits the divergences from each other. It suggests a stronger form the RTRD with the independent renormalization of each pack of lightlike Wilson lines that share the same transverse plane at light-cone infinity (see detailed description in sec.\ref{sec:geom_rap_div}). The rigorous proof of this stronger form of RTRD requires the demonstration of iterative subtraction for rapidity divergences. We expect that it can be performed with the help of Ward identities for the rapidity divergent contributions. Nonetheless, we were not able to pass through this procedure, because in order to disentangle different rapidity divergences a special (singular) gauge fixation condition should be used, which greatly complicates the task. 

The formulation of RTRD is made at a finite (although arbitrary) perturbative order. On  one hand, it is a consequence of necessity to use the perturbation theory to pass from QCD at the critical coupling to the physical coupling. On another hand sending the order to infinity and studying the asymptote of the perturbation expansions one recovers a part of the non-perturbation corrections associated with renomalon contributions. Therefore, we expect that RTRD can be used non-perturbatively at least for the renormalon contributions. The explicit leading order evaluation confirms it \cite{Scimemi:2016ffw}.

We have derived the all-order color structure of the MPS soft factor and presented its decomposition (up to three-loop order inclusively) in the terms of the generating function. In this way, we have checked the equivalence of the color structure of the soft anomalous dimension (SAD) and the rapidity anomalous dimension (RAD), which is predicted by RTRD. In turn the simple structure of MPS soft factor results to all-order constraints on the SAD. Namely, it predicts the absence of odd-color contributions at all orders, which is not known to our best knowledge. We have also presented in details the SAD-to-RAD correspondence discovered in \cite{Vladimirov:2016dll}, which predicts the three-loop expression for RAD using the finite-$\epsilon$ two-loop calculation \cite{Echevarria:2015byo}, and the three-loop expression for SAD \cite{Moch:2004pa,Almelid:2015jia}. The obtained three-loop RAD coincides with the calculation made in \cite{Li:2016ctv} by bootstrapping the decompositions of TMD and fully differential soft factors. This agreement shows a non-trivial confirmation of RTRD.

\acknowledgments The author gratefully acknowledges V.Braun, A.Manashov, and I.Scimemi for numerous stimulating discussions, and M.Diehl for important comments and help with the definition of multi-parton distributions.

\appendix

\section{$\delta$-regularization}
\label{app:delta-reg}

The general part of the discussion presented in the article is not restricted to any regularization procedure. For the examples we use the $\delta$-regularization. The connection between $\delta$-regularization and the regularization by the tilted Wilson lines can be found in \cite{Buffing:2017mqm} (see Appendix B).   

The $\delta$-regularization has been consistently formulated in \cite{Echevarria:2015byo}, and used in NNLO calculation \cite{Echevarria:2015usa,Echevarria:2016scs,Vladimirov:2016qkd}. The $\delta$-regularization consists in the following modification of the Wilson line
\begin{eqnarray}
\mathbf{\Phi}_v(x)\Big|_{\delta-\text{reg.}}=P\exp\(ig \int_0^\infty d\sigma v^\mu A_\mu^A(v\sigma+x)\mathbf{T}^A e^{-|(v\cdot \delta)|\sigma}\).
\end{eqnarray}
The $\delta$-regularization completely regularizes the rapidity divergences and IR-divergences associated with Wilson lines. To regularize the UV divergences the dimensional regularization is used with $d=4-2\epsilon$ (with $\epsilon>0$). 

The $\delta$-regularization is convenient for practical evaluation. The first, it gives a clear separation of rapidity and IR divergences. The rapidity divergences arise as a logarithms of $\delta$. The IR-divergences arise as $\epsilon-$power of $\delta$, e.g. $(\delta^+\delta^-)^{-\epsilon}$. Since $\epsilon>0$, such contribution is explicitly singular. Note, that this separation is clear only at non-zero $\epsilon$. Therefore, we demand that the limit $\delta\to 0$ is taken prior to $\epsilon\to 0$. However, this demand is not necessary for IR-safe matrix-elements. The second, the $\delta$-regularization is defined as a modification of Wilson line operator (in contrast to regularizations which modify e.g. the loop-integral measure). Therefore, the $\delta$-regularization can be applied to any configuration of Wilson lines. At last, the loop calculus with the $\delta$-regularization is simple, due to the fact that it preserves the lightlike vectors.

The IR and rapidity divergences are clearly distinguished within the $\delta$-regularization. Let us demonstrate it for the generic one-loop integral $I_{ij}$ given in (\ref{rapdiv:1-loop}). In the $\delta$-regularization the integral reads
\begin{eqnarray}
I_{ij}&=&2^{2-2\epsilon}\Gamma(1-\epsilon) \int_0^\infty d L
\frac{2L(v_i\cdot v_j)}{(2(v_i\cdot v_j)L^2+b_{ij}^2+i0)^{1-\epsilon}}\int_0^\infty \frac{d\alpha}{\alpha}e^{-L \delta_i\alpha }e^{-L\delta_j/\alpha },
\end{eqnarray}
where $\delta_i=(v_i\cdot \delta)$, and $b^2_{ij}=-(b_i-b_j)^2>0$. 
The rapidity-divergent regimes $\alpha\to 0$ or $\alpha \to \infty$ result into $\ln\delta_j$ and $\ln\delta_j$ correspondingly. In the IR-regime then $L\to \infty$ the integral has only single dimensional parameter $\delta^2=2\delta_i\delta_j$, and therefore is proportional to $(\delta^2)^{-\epsilon}$. This contribution is singular at $\epsilon>0$ and $\delta \to 0$, and represents the IR-singularity. Indeed, evaluating the integral $I_{ij}$ we obtain
\begin{eqnarray}
I_{ij}&=&2\Gamma^2(\epsilon)\Gamma(1-\epsilon)\(\frac{2\delta_i \delta_j}{(v_i\cdot v_j)}\)^{-\epsilon}-2 \Gamma(-\epsilon)\(\frac{b_{ij}^2}{4}\)^\epsilon
\(\ln\(\frac{b_{ij}^2}{4}\frac{2\delta_i \delta_j}{(v_i\cdot v_j)}\)-\psi(-\epsilon)+\gamma_E\).
\end{eqnarray}
Such structure holds for arbitrary difficult loop integral, due to the fact that rapidity divergences insensitive to the dimensional regularization, while IR-divergences should be regularized at $\epsilon<0$.

The negative point of the $\delta$-regularization is the violation of the gauge-transformation properties of the Wilson line. However, these contributions are easy to trace, since gauge violating contributions are given by the positive powers of $\delta$. Therefore, in the calculation one should keep the parameter $\delta$ infinitesimal\footnote{One should pay special attention to the power-like IR divergences, e.g. $\delta^{-1-\epsilon}$. These divergences can interfere with the higher-order terms in the $\delta$-expansion and compensate each other. This case leads to the gauge violating contributions. However, these divergences are simple to track. See detailed discussion is given in the appendix of \cite{Echevarria:2015byo}.}, which makes loop-calculus even simpler. 

\subsection{Cancellation of mass-divergences in $\delta$-regularization}
\label{sec:delta-structure}

Any $n$-loop diagram contributing to the MPS soft factor in the $\delta$-regularization has a generic form
\begin{eqnarray}\label{rad:gen_integral}
\mathbf{M}^{[n]}=(\delta^2)^{-n\epsilon}\mathbf{A}^{[n]}_n(\epsilon)+(\delta^2)^{-(n-1)\epsilon}(b^2)^{\epsilon}\mathbf{A}^{[n]}_{n-1}(\ln(\delta^2),\{b\},\epsilon)+...+
(b^2)^{n\epsilon}\mathbf{A}^{[n]}_{0}(\ln(\delta^2),\{b\},\epsilon),
\end{eqnarray}
where $b^2$ is a transverse distance, say $b^2=(b_1-b_2)^2$, and $\mathbf{A}$ are dimensionless functions of transverse distances, rapidity divergent logarithms and parameter $\epsilon$. Note, that due to the Lorentz invariance the regularization parameters $\delta_{i,j}$ can appear only the combination $\delta^2$.

If color indices of MPS form a singlet, the IR-divergences cancel at each order of perturbation theory. It can be proven as following. Let us rescale $b_i\to l b_i$. If the color indices form singlets, the MPS soft factor should reduce to unity in the limit $\lambda\to 0$,
\begin{eqnarray}\label{rap:to_unity}
\lim_{\lambda\to 0}\mathbf{\Sigma}(\{\lambda b\},\delta)=\mathbf{I}.
\end{eqnarray}
It is the consequence of operator identity $\mathbf{\Phi}^\dagger_v(z)\mathbf{\Phi}_v(z)=\mathbf{I}$, which holds at arbitrary $\delta$ (even not infinitesimal). Therefore, the sum over diagrams at any given order vanishes in this limit
\begin{eqnarray}\label{rap:11}
\lim_{\lambda\to 0}\sum_{\text{diag.}}\mathbf{M}^{[n]}(\{\lambda b\},\delta)=0.
\end{eqnarray}
The functions $\mathbf{A}$ being dimensionless dependent on $\lambda$ only logarithmically. Therefore, all entires $\mathbf{A}_{i\neq n}$ in the expression (\ref{rad:gen_integral}) vanish in the limit $\lambda\to0$. Consequently, we have
\begin{eqnarray}\label{rap:22}
\sum_{\text{diag.}}\mathbf{A}^{[n]}_n(\epsilon)=0.
\end{eqnarray}
Next, we rescale $\delta^2\to \delta^2 \lambda^{2/(n-1)}$. The relation (\ref{rap:to_unity}) holds. Considering (\ref{rap:11}) we obtain $\sum_{\text{diag.}}\mathbf{A}^{[n]}_{n-1}(\epsilon)=0.$ On the next step we rescale $\delta^2\to \delta^2 l^{2/(n-2)}$, and demonstrate the absence of $\mathbf{A}^{[n]}_{n-2}$ constitutions. And so on. In this way, we obtain that
\begin{eqnarray}
\sum_{\text{diag.}}\mathbf{A}^{[n]}_k(\epsilon)=0,\qquad k>0.
\end{eqnarray}

The cancellation of IR divergences takes a place \textit{only} for color-singlet components of the MPS soft factor. The colored contributions are IR divergent, which can be seen already at NLO (see e.g.(\ref{app:MPS_colored})). In the spirit of presented discussion, the colored contributions do not obey the relation (\ref{rap:to_unity}), and thus, should not cancel in the sum of diagrams. Practically, it is convenient to keep contributions $\mathbf{A}_{k>0}^{[n]}$ in the diagrams, since they cancellation presents a nice check of the calculation.

\section{Generating function decomposition of MPS soft factor}
\label{app:color}

The generating function approach for the exponentiation of matrix elements of Wilson lines has been elaborated in \cite{Vladimirov:2015fea,Vladimirov:2014wga}. It naturally generalizes the well-known non-Abelian exponentiation technique for Wilson loops \cite{Gatheral:1983cz,Frenkel:1984pz}, onto the arbitrary configuration of Wilson lines. It is a powerful method which decouples the external color structure (i.e. the color part related to the Wilson lines, but not to the intrinsic loops) from the momentum integration. In this approach the final expression is given in the term of color generators and generating functions: the connected matrix elements of operator $V$, which are discussed later.

The operators $V$ are by-products of Wilson lines, and inherit their geometrical structure. It is convenient to present the final result via generating functions $W$ defined on the most elementary geometrical structures. In the case of MPS soft factor these are straight lightlike ray or paths of individual $\Phi$'s. However, color indices are contracted between pairs of $\mathbf{\Phi}$'s and thus, from the point of color-decomposition, $\mathbf{\Phi}$ is not an elementary object. 

There are two principal approaches in this situation. The first approach is to decouple the color indices at the transverse plane. The resulting object $\widetilde{\mathbf{\Sigma}}$ has $2N$-pairs of color indices. It can be straightforwardly written in the terms of elementary generating functions $W$ as
\begin{eqnarray}
\widetilde{\mathbf{\Sigma}}=e^{\mathbf{T}^{A_1}_1\frac{\partial}{\partial \theta^{A_{1}}_1}}...e^{\mathbf{T}^{A_{2N}}_{2N}\frac{\partial}{\partial \theta^{A_{2N}}_{2N}}}e^{W[\theta]}\Big|_{\theta=0}.
\end{eqnarray}
The color indices are coupled within the differential operator, which produces a more complicated operator, that act on the generating exponent,
\begin{eqnarray}
\mathbf{\Sigma}=e^{\pmb{\mathcal{D}}[\frac{\partial}{\partial \theta_1},\frac{\partial}{\partial \theta_{N+1}}]}...
e^{\pmb{\mathcal{D}}[\frac{\partial}{\partial \theta_N},\frac{\partial}{\partial \theta_{2N}}]}e^{W[\theta]}\Big|_{\theta=0},
\end{eqnarray}
with 
\begin{eqnarray}\nn
\pmb{\mathcal{D}}[\frac{\partial}{\partial \theta_1},\frac{\partial}{\partial \theta_{N+1}}]=\ln\(e^{\mathbf{T}^A\frac{\partial}{\partial \theta^A_1}}e^{\mathbf{T}^B\frac{\partial}{\partial \theta^B_{N+1}}}\),
\end{eqnarray}
where matrices $\mathbf{T}$ are contracted. This approach has been used in \cite{Vladimirov:2016qkd} for the calculation of DPS soft factor.

The second approach applies the same procedures in the opposite order. The expression for the MPS soft factor reads
\begin{eqnarray}\label{app:formula1}
\mathbf{\Sigma}=e^{\mathbf{T}^{A_1}_1\frac{\partial}{\partial \theta^{A_{1}}_1}}...e^{\mathbf{T}^{A_{N}}_{N}\frac{\partial}{\partial \theta^{A_{N}}_{N}}}e^{\mathcal{W}[\theta]}\Big|_{\theta=0}=e^{\pmb{\mathcal{W}}+\pmb \delta \pmb{\mathcal{W}}[\mathcal{W}]},
\end{eqnarray}
where $\mathcal{W}$ is the generating functions $\mathcal{W}$ for operators $\mathcal{V}$, which are defined on the cusped paths, $\pmb{\mathcal{W}}=\mathcal{W}[\mathbf{T}]$, and $\pmb \delta \pmb{\mathcal{W}}$ is the algebraic function of $\mathcal{W}$. The function $\delta\mathcal{W}$ is derived and discussed in details in ref.\cite{Vladimirov:2014wga}, and is called the defect of exponential procedure. In the turn, the operators $\mathcal{V}$ defined on an arbitrary paths can be rewritten in the terms of operators $V$ defined on elementary segments. Consequently, the generating function $\mathcal{W}$ can be presented in the terms of elementary generating functions $W$, and substituted to (\ref{app:formula1}).

In the following, we present in the details the calculation performed within the second approach.

\subsection{Evaluation of color structure}

The MPS soft factor given in eq.(\ref{SF:MPS}) can be conveniently presented in the form
\begin{eqnarray}\label{app:MPS_SF}
\mathbf{\Sigma}(b_1,b_2,...,b_N)=\langle 0|T\{\mathbf{\Lambda}(b_N)...\mathbf{\Lambda}(b_2)\mathbf{\Lambda}(b_1)\}|0\rangle,
\end{eqnarray}
where $\mathbf{\Lambda}(z)$ is a single Wilson lines build from two segment that meet at the point $z$,
\begin{eqnarray}
\mathbf{\Lambda}(z)=\mathbf{\Phi}_{-n}(z)\mathbf{\Phi}^\dagger_{-\bar n}(z).
\end{eqnarray}
The color indices are contracted at the cusp, but remain open on the ends of Wilson lines. 

The operator $\mathbf{\Lambda}$ can be written as 
\begin{eqnarray}
\mathbf{\Lambda}(z)=P\exp\(ig\int_{\gamma} dy^\mu A_\mu^A(y+z)\mathbf{T}^A\)=e^{\mathbf{T}^A \mathcal{V}_A},
\end{eqnarray}
where $\gamma$ is the path of Wilson line. The expression for the operators $\mathcal{V}$ can be found in \cite{Vladimirov:2015fea,Vladimirov:2014wga}. In the terms of generating functions for these operator the MPS soft factors takes the form
\begin{eqnarray}
\mathbf{\Sigma}(\{b\})=\(\prod_{i=1}^N e^{\mathbf{T}^{A_i}\frac{\partial}{\partial \theta^{A_i}_i}}\)e^{\mathcal{W}[\{\theta\},\{b\}]}\Big|_{\theta=0},
\end{eqnarray}
where 
\begin{eqnarray}
e^{\mathcal{W}[\{\theta\},\{b\}]}=\langle 0|e^{\sum_{i=1}^N \theta_i^{A}\mathcal{V}_i}|0\rangle,
\end{eqnarray}
where $\mathcal{V}_i=\mathcal{V}(b_i)$.

The generating function $\mathcal{W}$ contains only fully connected matrix element of various compositions of operators $\mathcal{V}$. It has the general form
\begin{eqnarray}\label{app:W[theta]}
\mathcal{W}[\theta]&=&\sum_i \theta_i^A \mathcal{W}_i^A+\frac{1}{2}\sum_{i,j} \theta_i^A \theta_j^B \mathcal{W}^{AB}_{ij}
+...+\frac{1}{n!}\sum_{i,j,..,k} \theta_i^A \theta_j^B..\theta_k^C \mathcal{W}^{AB..C}_{ij..k}+...~,
\end{eqnarray}
where the summation runs from $1$ to $N$ for each summation label, and
\begin{eqnarray}\label{app:calW_def}
\mathcal{W}^{AB..C}_{ij..k}=\langle\langle \mathcal{V}_i^A\mathcal{V}_j^B...\mathcal{V}_k^C \rangle\rangle,
\end{eqnarray}
where double brackets $\langle\langle..\rangle\rangle$ denote the connected part of the matrix element. Accordingly, $\mathcal{W}_{ij..}$ depends only on $\{b_i,b_j,...\}$. The functions $\mathcal{W}^{A..B}_{i..j}$ are necessarily symmetric over the permutations over the pairs of indices $(A,i)$. The matrix element $\langle\langle \mathcal{V}_1...\mathcal{V}_n\rangle\rangle$ has is proportional to $a_s^{n-1}$ at least.

The color indices of a generating function $\mathcal{W}^{A...B}_{i...j}$ are restricted to color-singlets, due to the global color-conservation. For the consideration of $a_s^3$-order the following function are required,
\begin{eqnarray}
\mathcal{W}^A_i&=&0,\nn
\\
\mathcal{W}^{AB}_{ij}&=&a_s\delta^{AB} \mathcal{W}_{ij},\label{app:W->Wsinglet}
\\
\mathcal{W}^{ABC}_{ijk}&=&a_s^2if^{ABC} \mathcal{W}_{ijk}+a_s^3 d^{ABC}\mathcal{W}^{(s)}_{ijk},\nn
\\
\mathcal{W}^{ABCD}_{ijkl}&=&a_s^3 \(if^{AB;CD}W_{ijkl}+if^{AC;BD}W_{ikjl}+if^{AD;BC}W_{iljk}\)+...,\nn
\end{eqnarray}
where $if^{AB;CD}=if^{AB\alpha}if^{\alpha CD}$, $d^{ABC}=2\Tr(\mathbf{T}_{adj}^A\{\mathbf{T}_{adj}^B\mathbf{T}_{adj}^C\})$. We have extracted the minimal perturbative order from functions $\mathcal{W}_{i...j}$, however, note that functions $\mathcal{W}_{i...j}$ have all perturbative orders. The dots in the last line of (\ref{app:W->Wsinglet}) denote the contributions of order $a_s^4$ which are accompanied by different color structures, e.g. by $f^{AB\alpha}d^{\alpha CD}$.

The action of the derivative exponent can be presented as the sum of terms (\ref{app:formula1}). The first term $\pmb{\mathcal{W}}$ is obtained from (\ref{app:W[theta]}) by substitution of sources $\theta^A_i$ by $\mathbf{T}^A_i$. The result of substitution can be written in the form
\begin{eqnarray}\label{app:W_eq2}
\pmb{\mathcal{W}}&=&\frac{a_s}{2}\sum_i C_{i}\mathcal{W}_{ii}+\frac{a_s}{2}\sum_{[i,j]}\mathbf{T}^A_i\mathbf{T}^A_j\mathcal{W}_{ij}
+\frac{a_s^2}{3!}\sum_{[i,j,k]}\mathbf{T}_i^A\mathbf{T}_j^B\mathbf{T}_k^C if^{ABC}\mathcal{W}_{ijk}
\\
&&\nn +\frac{a_s^3}{3!}\sum_{[i,j,k]}\mathbf{T}_i^A\mathbf{T}_j^B\mathbf{T}_k^C d^{ABC}\mathcal{W}_{ijk}^{(s)}
+a_s^3\sum_{[i,j]}\mathbf{T}_i^{\{AB\}}\mathbf{T}_j^C d^{ABC}\frac{\mathcal{W}_{iij}^{(s)}}{2}
+a_s^3\sum_{i}\mathbf{T}_i^{\{ABC\}}d^{ABC}\mathcal{W}_{iii}^{(s)}
\\
&&\nn +a_s^3\sum_{[i,j]}\mathbf{T}_i^{\{AB\}}\mathbf{T}_j^{\{CD\}}if^{AC;BD}\frac{\mathcal{W}_{ijij}}{4}
+a_s^3\sum_{[i,j,k]}\mathbf{T}_i^{\{AB\}}\mathbf{T}^C_j\mathbf{T}_k^D if^{AC;BD}\frac{\mathcal{W}_{ijik}}{2}
\\&&\nn
+a_s^3\sum_{[i,j,k,l]}\mathbf{T}_i^A\mathbf{T}_j^B\mathbf{T}_k^C\mathbf{T}_l^Dif^{AC;BD}\frac{\mathcal{W}_{ijkl}}{4}+\mathcal{O}(a_s^4),
\end{eqnarray}
where $C_i$ is the quadratic Casimir eigenvalue of $i$'th representation $C_i=\mathbf{T}_i^A\mathbf{T}_i^A$, and the summations run from $1$ to $N$ for each summation label, and none of labels are equal (which we denote by square brackets). The symmetric combinations of generators are labeled by curly brackets, $\mathbf{T}^{\{AB\}}_i=(\mathbf{T}^A_i\mathbf{T}^B_i+\mathbf{T}^B_i\mathbf{T}^A_i)/2$, $\mathbf{T}^{\{ABC\}}_i=(\mathbf{T}^A_i\mathbf{T}^B_i\mathbf{T}^C_i+...+\mathbf{T}^C_i\mathbf{T}^B_i\mathbf{T}^A_i)/6$, etc. To present the expression (\ref{app:W_eq2}) in compact form, Jacobi identities have been used.

The derivation of the general form for the defect contribution in given in ref.\cite{Vladimirov:2015fea}. It can be presented as $\pmb{\delta \mathcal{W}}=\pmb{\delta}_2 \pmb{\mathcal{W}}+\pmb{\delta}_3 \pmb{\mathcal{W}}+...$, where $\pmb{\delta}_n \pmb{\mathcal{W}}$ contains algebraic combinations of $n$ entries of $\mathcal{W}$. At $a_s^3$-order only two leading terms contribute. They are
\begin{eqnarray}
\pmb{\delta}_2 \pmb{\mathcal{W}}&=&\frac{\{\pmb{\mathcal{W}}^2\}}{2}-\frac{\pmb{\mathcal{W}}^2}{2},
\\
\pmb{\delta}_3 \pmb{\mathcal{W}}&=&\frac{\{\pmb{\mathcal{W}}^3\}}{6}-\frac{\pmb{\delta}_2 \pmb{\mathcal{W}}\,\pmb{\mathcal{W}}
+\pmb{\mathcal{W}}\,\pmb{\delta}_2 \pmb{\mathcal{W}}}{2}-\frac{\pmb{\mathcal{W}}^3}{6},
\end{eqnarray}
where curly brackets denote the complete symmetrization of generators. Evaluation of these expressions is straightforward. The results are
\begin{eqnarray}
\pmb{\delta}_2\pmb{\mathcal{W}}&=&-a_s^2\frac{C_A}{48}\sum_{i}C_i\mathcal{W}_{ii}^2+a_s^2\frac{C_A}{48}\sum_{[i,j]}\mathbf{T}_i^A\mathbf{T}_j^A\(
3\mathcal{W}_{ij}^2-4\mathcal{W}_{ij}\mathcal{W}_{ii}\)
\\\nn&&+
a_s^3\frac{C_A}{48}\sum_{[i,j,k]}\mathbf{T}_i^A\mathbf{T}_j^B\mathbf{T}_k^C if^{ABC}\mathcal{W}_{ijk}\(3\mathcal{W}_{ij}
-2\mathcal{W}_{ii}\)+\mathcal{O}(a_s^4),
\end{eqnarray}
\begin{eqnarray}
\pmb{\delta}_3\pmb{\mathcal{W}}&=&a_s^3 \frac{C_A^2}{576}\sum_i C_i \mathcal{W}_{ii}^3
\\\nn&&+a_s^3\frac{C_A^2}{576}\sum_{[i,j]}\mathbf{T}_i^A\mathbf{T}_j^A
\(
6\mathcal{W}_{ii}^2\mathcal{W}_{ij}-12\mathcal{W}_{ii}\mathcal{W}^2_{ij}+2\mathcal{W}_{ii}\mathcal{W}_{jj}\mathcal{W}_{ij}+5\mathcal{W}_{ij}^3
\)
\\\nn&&
+\frac{a_s^3}{24}\sum_{[i,j]}\mathbf{T}^{\{AB\}}_i\mathbf{T}^{\{CD\}}_jif^{AC;BD}\(\mathcal{W}_{ij}^3-\mathcal{W}_{ij}^2\mathcal{W}_{ii}\)
\\\nn&&
+\frac{a_s^3}{24}\sum_{[i,j,k]}\mathbf{T}_i^{\{AB\}}\mathbf{T}_j^C\mathbf{T}_k^D
if^{AC;BD}\(-\mathcal{W}_{ij}\mathcal{W}_{ik}\mathcal{W}_{jk}+2\mathcal{W}_{ij}\mathcal{W}_{ik}^2-\mathcal{W}_{ii}\mathcal{W}_{ij}\mathcal{W}_{ik}\)
\\\nn&&
+\frac{a_s^3}{24}\sum_{[i,j,k,l]}\mathbf{T}_i^{A}\mathbf{T}_j^B\mathbf{T}_k^C\mathbf{T}_l^D
if^{AC;BD}\(-\mathcal{W}_{ij}\mathcal{W}_{ik}\mathcal{W}_{jl}\)+\mathcal{O}(a_s^4),
\end{eqnarray}
where $C_A=N_c$.

\subsection{Segment reduction of the generating function}
\label{app:segment_decomposition}

To reduce the generating function $\mathcal{W}$ (that are connected matrix elements of $\mathcal{V}$) to elementary generating functions $W$ (that are connected matrix elements of $V$) we recall that
\begin{eqnarray}\label{app:eV=eVeV}
\mathbf{\Lambda}(z)=e^{\mathbf{T}^A \mathcal{V}_A(z)}=\mathbf{\Phi}_{-n}(z)\mathbf{\Phi}^\dagger_{-\bar n}=e^{-\mathbf{T}^A V^{n}_A(z)}e^{\mathbf{T}^B V^{\bar n}_B(z)}.
\end{eqnarray}
The different signs infront of $V^n$ and $V^{\bar n}$ are consequence of the definition of $V$ on the path from $0$ to infinity. Using Baker-Campbell-Hausdorff formula we obtain
\begin{eqnarray}
\mathcal{V}_A&=&-V^n_A+V^{\bar n}_A-\frac{if^{ABC}}{2}V^n_BV^{\bar n}_C+\frac{if^{AB;CD}}{12}\(
V^n_BV^n_CV^{\bar n}_D-V^{\bar n}_BV^{\bar n}_CV^n_D\)
\\\nn &&
-\frac{if^{AB;C;DE}}{24}V^{\bar n}_BV^n_CV^n_DV^{\bar n}_E+
\frac{if^{AB;C;D;EF}}{720}\Big(V^{\bar n}_BV^{\bar n}_CV^{\bar n}_DV^{\bar n}_EV^n_F-V^n_BV^n_CV^n_DV^n_EV^{\bar n}_F
\\&&\nn +2	V^n_BV^{\bar n}_CV^{\bar n}_DV^{\bar n}_EV^n_F-2V^{\bar n}_BV^n_CV^n_DV^n_EV^{\bar n}_F
 +6 V^{\bar n}_BV^n_CV^{\bar n}_DV^n_EV^{\bar n}_F-6V^n_BV^{\bar n}_CV^n_DV^{\bar n}_EV^n_F
\Big)
\\\nn &&+\mathcal{O}(g^6),
\end{eqnarray}
where we omit the arguments $z$  and
$$
if^{AB;C;...;EF}=if^{AB\alpha}if^{\alpha C\beta}if^{\beta...}...if^{...\gamma}if^{\gamma EF}.
$$

There is an important consequence of (\ref{app:eV=eVeV}), which gives exact restrictions on generating functions. We observe that
\begin{eqnarray}
\mathcal{V}_A=-\mathcal{V}_A(n\leftrightarrow \bar n).
\end{eqnarray}
Alternatively, the exchange of $n$ and $\bar n$ can done by a rotation. It implies
\begin{eqnarray}
\mathcal{W}=\mathcal{W}(n \leftrightarrow \bar n).
\end{eqnarray}
Therefore, the generating functions of odd power of $\mathcal{V}$ are exactly zero:
\begin{eqnarray}\label{app:odd=0}
\mathcal{W}_{i_1...i_{2n+1}}^{A_1...A_{2n+1}}=0.
\end{eqnarray}
For the case of generating functions $\mathcal{W}_{ijk}$ this statement was demonstrated in \cite{Vladimirov:2016qkd}. Important to note that absence  of generating functions with odd-number of indices does not imply the absence of diagrams which connect odd number of Wilson lines. Such contributions are possible, but the number of connections will be even. The generators belonging to the same Wilson lines always appear in the symmetric composition.  The general all-order structure can be written as
\begin{eqnarray}\label{app:color-hyirarchy}
\mathbf{\Sigma}(\{b\})=\exp\(\sum_{\substack{n=2\\n\in\text{even}}}^{\infty}a_s^{n/2}\sum_{i_1,...,i_n=1}^N\{\mathbf{T}^{A_1}_{i_1}...\mathbf{T}^{A_n}_{i_n}\}\sigma^{n;i_1...i_n}_{A_1...A_n}(\{b\})\).
\end{eqnarray}
In the next paragraph we present the first two entries of this expression.

The $\mathcal{W}$ is given by the connected matrix element of $\mathcal{V}$'s, see (\ref{app:calW_def}). However, the operators $V$ inside $\mathcal{V}$ are not necessary connected. Therefore, we have to decompose matrix elements over their connected parts, e.g.
$$
\langle\langle \{V_1\}\{V_2V_3V_4\}\rangle\rangle=\langle\langle V_1V_2V_3V_4\rangle\rangle+
\langle\langle V_1V_2\rangle\rangle\langle\langle V_3V_4\rangle\rangle
+\langle\langle V_1V_3\rangle\rangle\langle\langle V_2V_4\rangle\rangle
+\langle\langle V_1V_4\rangle\rangle\langle\langle V_2V_3\rangle\rangle.
$$
In this example, $\{V_1\}$ and $\{V_2V_3V_4\}$ are resulted from separate $\mathcal{V}$'s and thus the connectivity between these operators should be preserved. By simple algebraic manipulations we arrive to the expressions for $\mathcal{W}$ in the terms of elementary generating functions. To present the result in the compact form let us introduce the notation which respect the symmetries of matrix elements
\begin{eqnarray}\nn
&&\langle\langle V_A^{v_i}(b_i)V_B^{v_j}(b_j) \rangle\rangle=a_s \delta_{AB}(v_i\cdot v_j)W(b_{ij}),
\\\label{app:W_def}
&&\langle\langle V_A^{v_i}(b_i)V_B^{v_j}(b_j)V_C^{v_k}(b_k) \rangle\rangle =a_s^2 i f_{ABC}\Big[
(v_i\cdot v_j)(v_i\cdot v_k)W(b_{ij},b_{ik},b_{jk})
\\\nn&&\qquad\qquad\qquad\qquad\qquad\qquad
+(v_i\cdot v_j)(v_j\cdot v_k)W(b_{jk},b_{ij},b_{ik})
+(v_j\cdot v_k)(v_i\cdot v_k)W(b_{ik},b_{jk},b_{ij})\Big],
\end{eqnarray}
where $b_{ij}=b_i-b_j$ and
$$
W(x,y,z)=-W(y,x,z).
$$
In the parameterization (\ref{app:W_def}), we have taken into account that Wilson lines are lightlike. The parametrization of the generating function of the fourth order is cumbersome. Therefore, we simply denote
\begin{eqnarray}
&&\langle\langle V_A^{v_i}(b_i)V_B^{v_j}(b_j)V_C^{v_k}(b_k)V_D^{v_l}(b_l) \rangle\rangle
\\\nn&&\qquad\qquad=a_s^3\(i f^{AB;CD}W^{v_iv_jv_kv_l}_{ijkl}+if^{AC;BD}W^{v_iv_kv_jv_l}_{ikjl}+if^{AD;BC}W^{v_iv_lv_jv_k}_{iljk}\)\nn
+\mathcal{O}(a_s^4).
\end{eqnarray}
The generating functions $\mathcal{W}$ reads
\begin{eqnarray}
\mathcal{W}_{ij}&=&
-2W(b_{ij})+a_s C_A\(-2 W(b_{ij},0,b_{ij})+\frac{W^2(b_{ij})}{4}-\frac{W(b_{ij})W(0)}{3}\)
\\&&\nn
a_s^2 C_A^2 \Bigg(-\frac{W^3(b_{ij})}{48}+\frac{5 W^2(b_{ij})W(0)}{24}-\frac{7W(b_{ij})W^2(0)}{72}
\\&&\nn \frac{W_{iijj}^{n\bar n n\bar n}}{4}+\frac{W_{ijij}^{n n \bar n \bar n}+W_{ijij}^{n \bar n n \bar n}}{8}
+\frac{W_{ijjj}^{n n n \bar n}+W_{ijjj}^{n\bar n n\bar n}
+W_{jiii}^{n n n \bar n}+W_{jiii}^{n\bar n n\bar n}}{8}\Bigg)+\mathcal{O}(a_s),
\end{eqnarray}
\begin{eqnarray}\nn
\mathcal{W}_{ijkl}=2\Big(W_{ijkl}^{nn\bar n\bar n}+W_{ijkl}^{n\bar nn\bar n}+W_{ijkl}^{\bar nnn\bar n}
-W_{ijkl}^{nnn\bar n}- W_{ijkl}^{nn\bar n n}-W_{ijkl}^{n\bar nn n}-W_{ijkl}^{\bar nnnn}\Big)+\mathcal{O}(a_s).
\end{eqnarray}

\subsection{Result}
\label{app:results}
Finally, we combine the expression for the MPS soft factor in the form
\begin{eqnarray}\label{app:MPS_colored}
\mathbf{\Sigma}&=&\exp\Bigg[a_s\sum_i C_i X_{0}+a_s \sum_{[i,j]}\mathbf{T}_i^A\mathbf{T}_j^A X_{2}^{ij}+
a_s^3 \Big(\sum_{[i,j]}\mathbf{T}_i^{\{AB\}}\mathbf{T}_j^{\{CD\}}if^{AC;BD}X_{4}^{ij}
\\\nn &&+\sum_{[i,j,k]}\mathbf{T}_i^{\{AB\}}\mathbf{T}_j^C\mathbf{T}_k^D if^{AC;BD}X_{4}^{ijk}
+\sum_{[i,j,k,l]}\mathbf{T}_i^{A}\mathbf{T}_j^B\mathbf{T}_k^C\mathbf{T}_l^D if^{AC;BD}X_{4}^{ijkl}\Big)+\mathcal{O}(a_s^4)\Bigg].
\end{eqnarray}
The functions $X$ are
\begin{eqnarray}
X_{0}&=&-W(0)-a_s\frac{C_A}{8}W^2(0)+\frac{a_s^3 C_A^2}{16} \(\frac{7}{18}W^3(0)+5W_{iiii}^{n\bar nn\bar n}\)+\mathcal{O}(a_s^3),
\\
X_{2}^{ij}&=&-W(b_{ij})+\frac{a_s C_A}{2}\(\frac{3}{4}W^2(b_{ij})-W(b_{ij})W(0)-2 W(b_{ij},0,b_{ij})\)
\\\nn&&+\frac{a_s^2C_A^2}{16}\Big(-\frac{41}{18}W^3(b_{ij})+\frac{19}{3}W^2(b_{ij})W(0)-\frac{11}{3}W(b_{ij})W^2(0)
\\\nn&&+8W(b_{ij})W(b_{ij},0,b_{ij})
-\frac{16}{3}W(b_{ij})W(b_{ij},0,b_{ij})+
\\\nn&&
2W_{iijj}^{n\bar n n\bar n}+W_{ijij}^{n n \bar n \bar n}+W_{ijij}^{n \bar n n \bar n}
+2W_{ijjj}^{n n n \bar n}+2W_{ijjj}^{n\bar n n\bar n}\Bigg)+\mathcal{O}(a_s^3)
\\
X_{4}^{ij}&=&-\frac{W^3(b_{ij})}{3}+\frac{W^2(b_{ij})W(0)}{3}
\\\nn&&+\frac{W_{ijij}^{nn\bar n\bar n}+W_{ijij}^{n\bar n n\bar n}
+W_{ijij}^{\bar nnn\bar n}}{2}-W_{ijij}^{nnn\bar n}-W_{ijij}^{nn\bar n n}+\mathcal{O}(a_s)
\\
X_{4}^{ijk}&=&\frac{W(b_{ij})W(b_{ik})}{3}\(W(b_{jk})+W(0)-2W(b_{ik})\)+\mathcal{W}_{ijik}+\mathcal{O}(a_s),
\\
X_{4}^{ijkl}&=&\frac{W(b_{ij})W(b_{ik})W(b_{jl})}{3}+\frac{\mathcal{W}_{ijkl}}{4}+\mathcal{O}(a_s).
\end{eqnarray}
To present expression in this form, we have used the permutation symmetries, the fact that indices $i,j,k,l$ are summed, and the Jacobi identities.

If we impose the color conservation condition 
\begin{eqnarray}
\sum_{i=1}^N\mathbf{T}_i^A=0,
\end{eqnarray}
we can eliminate some terms in favor of another terms. E.g. the following decomposition looks reasonable
\begin{eqnarray}
\mathbf{\Sigma}&=&\exp\Bigg[-a_s \sum_{[i,j]}\mathbf{T}_i^A\mathbf{T}_j^A \sigma(b_{ij})
\\\nn&&
+a_s^3 \Big(\sum_{[i,j,k]}\mathbf{T}_i^{\{AB\}}\mathbf{T}_j^C\mathbf{T}_k^D if^{AC;BD}Y_{4}^{ijk}
+\sum_{[i,j,k,l]}\mathbf{T}_i^{A}\mathbf{T}_j^B\mathbf{T}_k^C\mathbf{T}_l^D if^{AC;BD}X_{4}^{ijkl}\Big)+\mathcal{O}(a_s^4)\Bigg].
\end{eqnarray}
where
\begin{eqnarray}
\sigma(b_{ij})&=&X_0-X_2^{ij}-\frac{C_A^2}{2}X_4^{ij},
\\
Y_4^{ijk}&=&X_4^{ijk}-\frac{X_4^{ij}+X_4^{ik}}{2}.
\end{eqnarray}
This decomposition can be written in the form used in \cite{Almelid:2015jia} as
\begin{eqnarray}
\mathbf{\Sigma}&=&\exp\Bigg[-2a_s \sum_{i<j}\mathbf{T}_i^A\mathbf{T}_j^A Y_{2}^{ij}
+a_s^3 \Big(2\sum_{\substack{j<k\\i\neq j,k}}\mathbf{T}_i^{\{AB\}}\mathbf{T}_j^C\mathbf{T}_k^D if^{AC;BD}Y_{4}^{ijk}
\\\nn&&
+4\sum_{i<j<k<l}\mathbf{T}_i^{A}\mathbf{T}_j^B\mathbf{T}_k^C\mathbf{T}_l^D 
\(if^{AB;CD}X_{4}^{ijkl}+if^{AC;BD}X_{4}^{ikjl}+if^{AD;BC}X_{4}^{iljk}\)\Big)+\mathcal{O}(a_s^4)\Bigg].
\end{eqnarray}

\section{Explicit expressions for anomalous dimensions}
\label{app:ADs}

In this appendix, we collect the expressions for anomalous dimensions which are used in the text. 

The QCD $\beta$ function and its leading coefficients are
\begin{eqnarray}
\beta(a_s)&=&\sum_{n=0}^\infty \beta_n a_s^{n+1},
\\\nn \beta_0&=& \frac{11}{3}C_A-\frac{2}{3}N_f,\qquad \beta_1= \frac{34}{3}C_A^2-\frac{10}{3}C_AN_f-2C_FN_f,
\end{eqnarray}
where $C_A=N_c$ and $C_F=(N_c^2-1)/2N_c$ are eigenvalues of quadratic Casimir operator for adjoint and fundamental representations, and $N_f$ is the number of flavors. The cusp-anomalous dimension and its leading coefficients \cite{Moch:2004pa} are
\begin{eqnarray}
\Gamma_{cusp}^i &=& 4 C_i\sum_{n=0}^\infty a_s^{n+1}\Gamma_n,
\\\nn
\Gamma_0&=&1,\qquad \Gamma_1=\(\frac{67}{9}-2\zeta_2\)C_A-\frac{10}{9}N_f,
\\\nn
\Gamma_2&=&C_A^2\(\frac{245}{6}-\frac{268}{9}\zeta_2+22\zeta_4+\frac{22}{3}\zeta_3\)+C_AN_f
\(-\frac{209}{27}+\frac{40}{9}\zeta_2-\frac{56}{3}\zeta_3\)
\\\nn&&\qquad\qquad\qquad\qquad\qquad\qquad+C_FN_f\(-\frac{55}{6}+8\zeta_3\)-\frac{4 N_f^2}{27},
\end{eqnarray}
where $i$ is a representation of Wilson lines. The non-cusp part of the SAD and its leading coefficients \cite{Moch:2004pa} are
\begin{eqnarray}
\tilde \gamma^i_s&=&C_i\sum_{n=0}^\infty a_s^{n+1}\gamma_n,
\\\nn  \gamma_0&=&0,\qquad 
\gamma_1=C_A\(-\frac{808}{27}+\frac{22}{3}\zeta_2+28 \zeta_3\)+\(\frac{112}{27}-\frac{4}{3}\zeta_2\)N_f
\\\nn
\gamma_2&=&C^2_A\(-\frac{136781}{729}+\frac{12650}{81}\zeta_2+\frac{1316}{3}\zeta_3-176\zeta_4-\frac{176}{3}\zeta_2\zeta_3-192\zeta_5\)
\\\nn&&+
\(\frac{11842}{729}-\frac{2828}{81}\zeta_2-\frac{728}{27}\zeta_3+48\zeta_4\)C_AN_f+\(\frac{1711}{27}-4\zeta_2-\frac{304}{9}\zeta_3-16\zeta_4\)C_FN_f
\\\nn&&+\(\frac{2080}{729}+\frac{40}{27}\zeta_2-\frac{112}{27}\zeta_3\)N_f^2.
\end{eqnarray}

The rapidity anomalous dimension has the form
\begin{eqnarray}\label{app:d}
\mathcal{D}^i&=&C_i \sum_{n=1}^\infty a_s^n \sum_{k=0}^n L_b^k d^{(n,k)},\qquad L_b=\ln\(\frac{\mu^2 b^2}{4 e^{-2\gamma_E}}\),
\\\nn
d^{(1,0)}&=&0,\qquad d^{(2,0)}=C_A\(\frac{404}{27}-14 \zeta_3\)-\frac{56}{27}N_f,
\\\nn
d^{(3,0)}&=&C_A^2\(\frac{297029}{1458}-\frac{3196}{81}\zeta_2-\frac{6164}{27}\zeta_3-\frac{77}{3}\zeta_4+\frac{88}{3}\zeta_2\zeta_3+96\zeta_5\)
\\\nn&&+C_AN_f\(-\frac{31313}{729}+\frac{412}{81}\zeta_2+\frac{452}{27}\zeta_3-\frac{10}{3}\zeta_4\)
\\\nn&&+C_FN_f\(-\frac{1711}{54}+\frac{152}{9}\zeta_3+8\zeta_4\)+N_f^2\(\frac{928}{729}+\frac{16}{9}\zeta_3\).
\end{eqnarray}
The coefficients $d^{(n,i>0)}$ can be expressed in the terms of other anomalous dimensions and are given in sec.\ref{sec:SADRAD}.

\bibliographystyle{JHEP}
\bibliography{RtRd}

\end{document}